\documentclass[twoside]{article}

\usepackage{arxiv}
\renewcommand{\undertitle}{}
\renewcommand{\headeright}{}

\usepackage{natbib}

\usepackage{amsmath, amssymb, graphicx, bm}
\usepackage{amsthm}

\newtheorem{theorem}{Theorem}[section]
\newtheorem{lemma}[theorem]{Lemma}

\usepackage{bbm}
\usepackage{hyperref}
\usepackage{xcolor}
\usepackage{float}
\usepackage{booktabs}
\usepackage{bbding}
\usepackage{makecell}
\usepackage{comment}

\usepackage{graphicx}
\usepackage[caption=false]{subfig}

\allowdisplaybreaks

\usepackage{macros}

\definecolor{cyan}{rgb}{0.0, 1.0, 1.0}
\definecolor{dodgerblue}{rgb}{0.12, 0.56, 1.0}
\definecolor{kellygreen}{rgb}{0.3, 0.73, 0.09}
\definecolor{jasper}{rgb}{0.84, 0.23, 0.24}
\definecolor{fandango}{rgb}{0.71, 0.2, 0.54}
\definecolor{amber}{rgb}{1.0, 0.75, 0.0}
\definecolor{indigo}{rgb}{0.29, 0.0, 0.51}

\newcounter{savedfigure}
\newcounter{algorithm}

\usepackage{bibunits}
\defaultbibliography{references}
\defaultbibliographystyle{apalike}

\usepackage{authblk}
\title{Testing Additivity of Lead and
Benzo[a]pyrene--induced Neurotoxicity \\ in \textit{Caenorhabditis elegans} Assays}
\author[1]{Niccol\`o Anceschi$^*$}
\author[2]{Javier Huayta}
\author[2]{Joel N.\ Meyer}
\author[1]{David B.\ Dunson}
\author[1]{Amy H.\ Herring}
\affil[1]{Department of Statistical Science, Duke University, Durham, NC}
\affil[2]{Nicholas School of the Environment, Duke University, Durham, NC}
\date{}
\renewcommand{\shorttitle}{Testing Toxicants Additivity in \textit{C. elegans} Neurotoxicity Assays}
\begin{document}

\maketitle
\renewcommand{\thefootnote}{$~$}
\footnotetext{{\normalsize $^*$} Corresponding author: \texttt{niccolo.anceschi@duke.edu}}
\renewcommand{\thefootnote}{\arabic{footnote}}

\begin{bibunit}

\vspace{-20pt}

\begin{abstract}
    Exposure to environmental contaminants is a recognized cause of neurotoxicity, contributing to the onset of a broad range of neurological conditions.
    In realistic settings, such exposure involves complex mixtures, and the combined effect of their components may differ from what their individual effects would predict.
    Characterizing such interactions and testing them against a principled notion of additivity is central to assessing the neurotoxicological risk.
    We take up these questions for two widespread and independently neurotoxic pollutants, lead (Pb) and benzo[a]pyrene (BaP), through a novel \textit{C. elegans} assay in which nematodes were subjected to single and joint exposures across a range of doses.
    Morphological damage is quantified on an ordinal scale at the level of individual dopaminergic neurons. 
    To analyze these data, we model the full distribution of the ordinal response as a convex mixture between an unexposed and a maximally affected profile.
    The weight of this mixture varies with chemical doses, modeled flexibly via monotone splines and, for the joint effect, in a radial coordinate system.
    Additivity is assessed via a likelihood ratio test against established null models, and is calibrated via parametric bootstrap.
    Applied to the \textit{C. elegans} assay, our analysis reveals a localized, asymmetric synergy between Pb and BaP, concentrated where moderate BaP meets high Pb exposure.
\end{abstract}


\section{Introduction}

Exposure to neurotoxic environmental compounds can progressively impair both the structure and the function of the nervous system, and the resulting degeneration is implicated in a broad range of neurological disorders. 
Clarifying how such exposures translate into neuronal damage is therefore an important step toward understanding the onset of many neurological conditions.
Exposure to toxicants rarely entails a single isolated compound, but rather mixtures of components whose joint action need not reduce to the individual contributions.
Assessing the combined effect of two or more substances requires a reference notion of \emph{additivity} -- the joint behavior expected in the absence of any interaction between them. 
Departures from this baseline are termed \emph{synergy} when the combined effect exceeds the additive expectation and \emph{antagonism} when it falls short. 
Quantifying such departures is central for a comprehensive assessment of toxicological risk.

\begin{figure}[ht!]
    \centering
    \includegraphics[width=\linewidth, trim=0 185 0 0, clip]{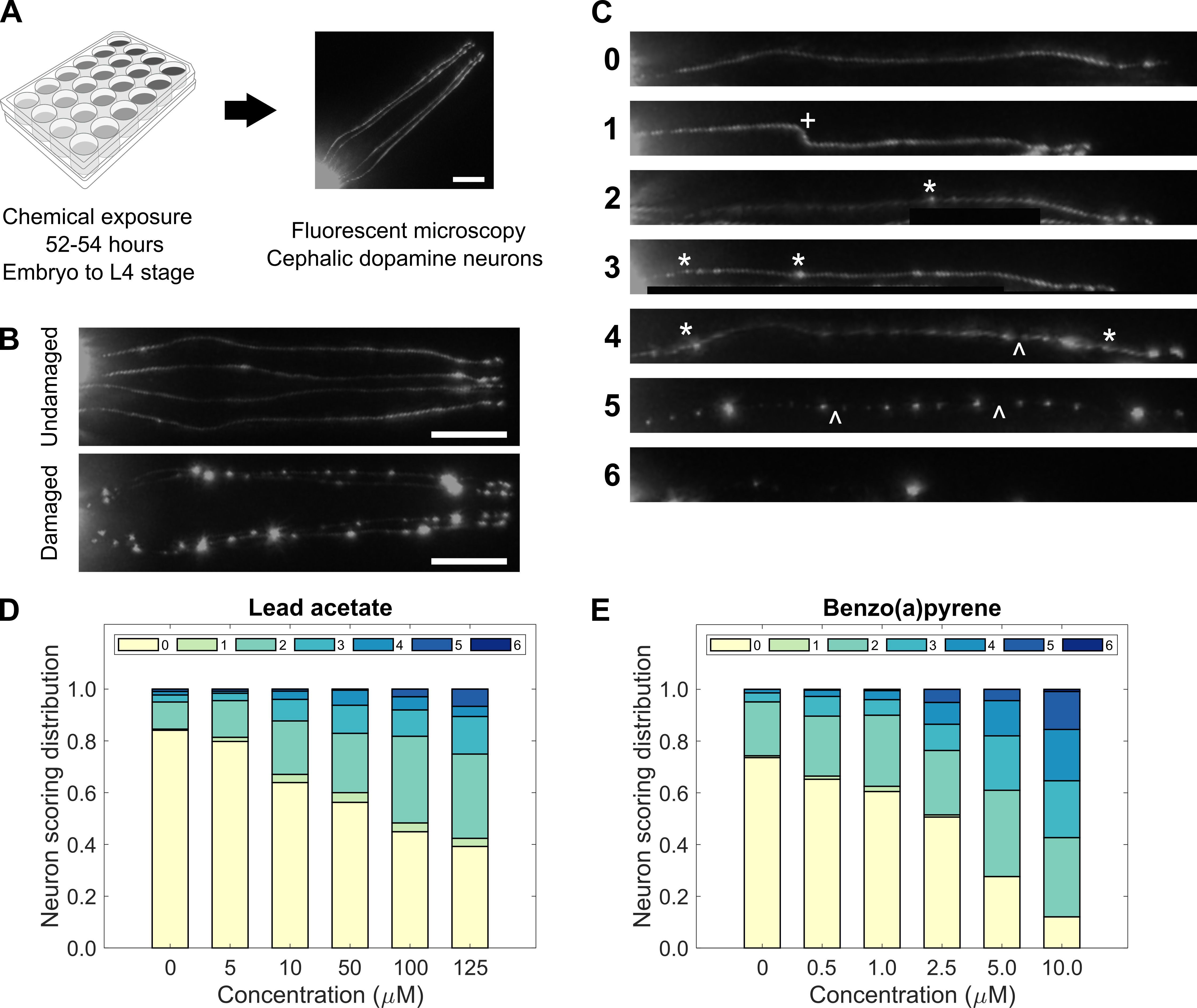}
    \vspace{-20pt}
    \caption{
    Representative images from the \textit{C. elegans} neurotoxicity assay and the ordinal damage scale of \citet{bijwadia2021quantifying}.
    (A) Experimental workflow: nematodes are exposed to chemicals from the embryo to the L4 stage (52–54 hours), after which cephalic dopaminergic neurons are imaged by fluorescence microscopy. 
    (B) Sample undamaged and damaged neurons, illustrating the morphological contrast the scale is built to capture.    
    Scale bars, [$X\sim$ um].
    (C) The ordinal scale, ranging from 0 (intact morphology) to 6 (complete loss of integrity), with damage features marked -- kinks or bends (+), blebs or swelling (*), and breaks or fragmentation (\^{}).
    }
    \vspace{-5pt}
    \label{fig_neuron_photos}
\end{figure}

We investigate this type of interaction between lead (Pb) and benzo[a]pyrene (BaP), two widespread neurotoxicants of considerable environmental and public health concern, which have been independently linked to dopaminergic neurodegeneration. 
{The two frequently co-occur in real-world exposure scenarios, from urban air pollution to industrial emissions and contaminated sites, yet regulatory thresholds are typically derived compound by compound \citep{USEPA1986_MixturesGuidelines, USEPA2000_MixturesGuidance}, and combined assessments default to additivity \citep{USEPA2023_DoseAddition}.
Characterizing whether their joint action departs from additivity is therefore directly relevant to the risk assessment of realistic co-exposures.
To this end, we analyze the combined action of Pb and BaP in novel experimental data, which use \textit{C. elegans} nematodes as a model organism.
% Our goal is to study the dose–response relationship linking Pb and BaP exposure to neuronal damage, and to test the additivity of their joint action. 
Crucially, toxicant-induced damage is recorded at the level of individual neurons on an ordinal severity scale, a data structure that most combination-analysis methods are not designed to handle. 
Representative neurons at each damage level are shown in Figure~\ref{fig_neuron_photos}, while a detailed description of the Toxicological assay is provided in Section~\ref{sec_data} and in Appendix~\ref{app_exp_protocol}.
}

\begin{figure}[H]
    \centering
    \includegraphics[width=\linewidth, trim=0 220 0 230, clip]{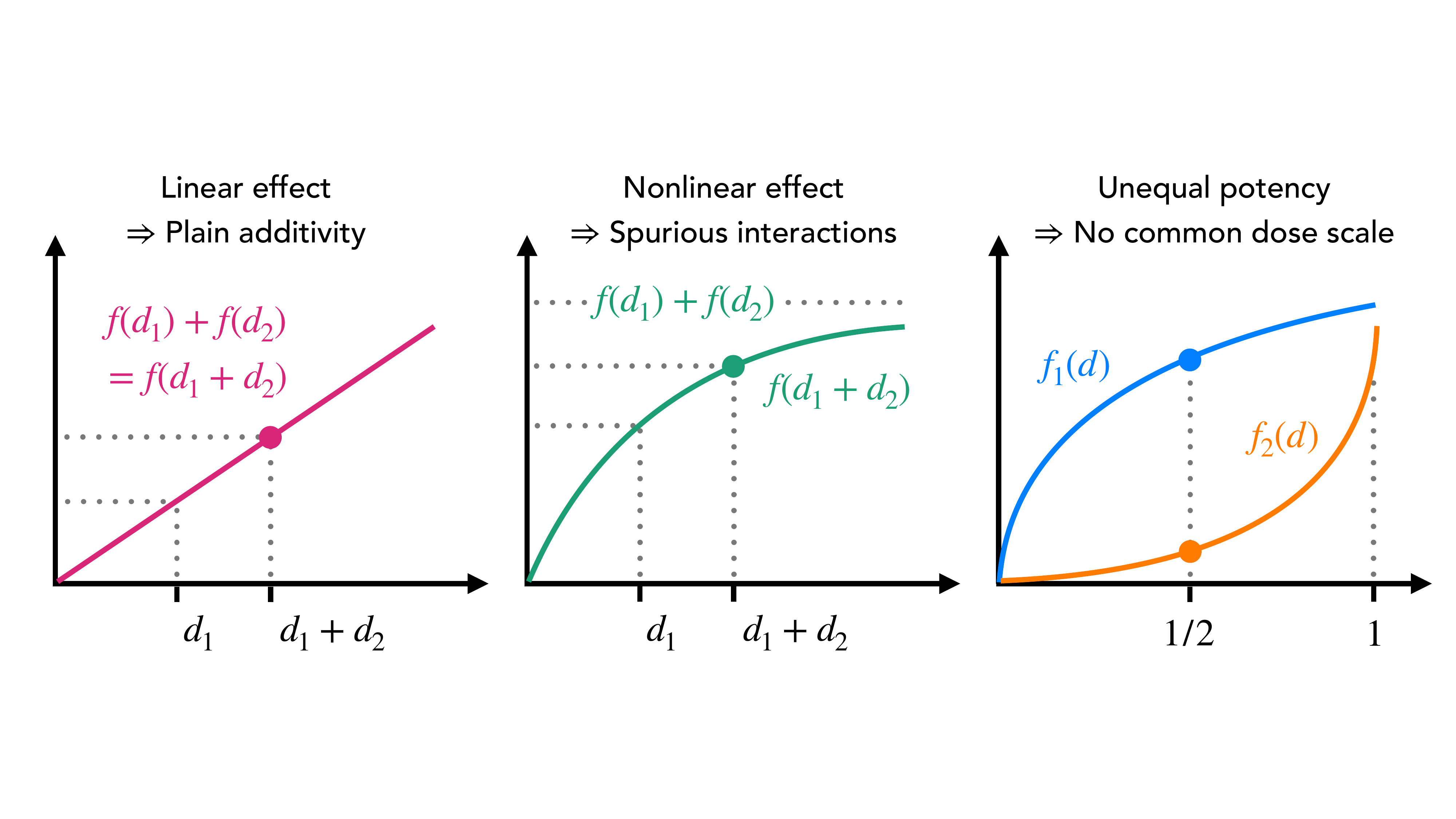}
    \put(-455,150){\makebox(0,0){\textbf{A}}}
    \put(-303,150){\makebox(0,0){\textbf{B}}}
    \put(-151,150){\makebox(0,0){\textbf{C}}}
    \vspace{-10pt}
    \caption{
    {
    \!\!\!\!\!
    Current statistical methods for modeling nonlinear interactions rely on a purely {algebraic notion of additivity} that lacks toxicological grounding. (A) Under the sham combination principle, the interaction term correctly identifies departures from additivity only when the single-exposure dose--response curve is linear, a condition that is rarely satisfied in practice. (B) For nonlinear dose--response curves, the interaction term would detect a spurious synergistic or antagonistic effect even for mechanistically identical chemicals, solely due to the nonlinearity of the dose--response relationship. (C) More generally, chemicals may differ in potency and dose--response shape. Even assuming the same dose range, direct dose summation would be ill-posed without first mapping doses into equivalent effects.}
    }
    \label{fig_interactions}
\end{figure}

{
A disconnect exists in how additivity is construed in the statistical literature versus the toxicological and pharmacological ones.
In statistical methodology, the analysis of chemical mixtures has largely concentrated on modeling synergy or antagonism through flexible yet parsimonious interaction terms \citep{Bobb2015_BKMR,Ferrari2020_interactions_GP,Chattopadhyay2025_SAID}.
Within a generalized-linear-model formulation, this reads
\begin{equation*}
\lambda^{-1}\!\big( \mathbb{E}[\, y \mid d_1, d_2 \,] \big) = f_1(d_1) + f_2(d_2) + \zeta(d_1,d_2) \;,
\end{equation*}
where $y$ is the response, $d_1$ are $d_2$ the two doses, $\lambda^{-1}(\cdot)$ is a link function, $f_1(\cdot)$ and $f_2(\cdot)$ capture the single-exposure effects, and $\zeta(\cdot\, , \cdot)$ the interaction among them.
While such models can achieve good fit, assessing departures from additivity by testing $\zeta (\cdot\, , \cdot) \neq0$ benchmarks against a toxicologically limited notion of additive action.
Consider two mechanistically identical chemicals: by the so-called \emph{sham combination principle}, no interaction should be present, and the model should collapse to
$$\lambda^{-1}\!\big( \mathbb{E}[ \, y \mid d_1, d_2 \,] \big) = f(d_1 + d_2) $$
for a common single-exposure curve $f(\cdot)$.
Instead, the additive-interaction form above would detect a spurious interaction $\zeta(\cdot\, , \cdot)\neq0$ whenever $f(\cdot)$ is nonlinear, as illustrated in Figure~\ref{fig_interactions}.
Since nonlinear effects are the norm rather than the exception, this would incorrectly flag synergy or antagonism where none exists.
}
\subsection{Combination Analysis: Background and Challenges}

{A toxicologically meaningful yet statistically rigorous} analysis of chemical combinations hinges on three interrelated methodological challenges:
(i) modeling the dose-response relationship for each substance individually,
(ii) combining them to specify a principled null model for additive joint action, 
and
(iii) constructing formal tests for departures from the chosen additive baseline.

The dose-response curve, quantifying how the outcome profile changes across exposure levels, is conventionally assumed to be monotonically non-decreasing \citep{hill2018nonmonotonic}.
Biologically, this is grounded in the expectation that cellular stress accumulates with increasing exposure, leading to more severe cellular damage.
For continuous and binary outcomes, a large body of pharmacological and toxicological
literature relies on sigmoidal Hill-type curves \citep{Yadav2015_ZIP,Twarog2016_Braid,Wooten2021_MuSyC}, or, more generally, on parametric models assuming linearity in log-dose after a suitable transformation \citep{Kong2006_loglinear,Demidenko2019Synergy}.
Although powerful when well-specified, 
these assumptions can be too rigid to capture empirically observed behavior,
leading to brittle inference under complex biological endpoints.
For ordered categorical responses, the situation is less developed.
Typical approaches either treat the outcome directly as continuous or collapse it into a naive summary statistic -- such as the mean -- that inevitably discards information.

\vspace{-5pt}

\vspace{-10pt}

Given individual dose-responses for each chemical, a central question in combination analysis is which null model best characterizes the baseline for additive joint action.
No general consensus exists in the literature \citep{Rider2018_ChemicalMixtures}, with different null models embodying distinct high-level mechanistic rationales, agnostic to the specific biological pathways or molecular targets involved.
Adjudicating between these nulls 
via simulation is complicated by the absence of an explicit model for the underlying biological processes.
Some contributions suggest sidestepping the choice of a single null, but rather recommend running multiple reference models and compounding the results \citep{DiVeroli2016_Combenefit,Ianevski2017_SynergyFinder,Ianevski2020_SynergyFinder2}, or assessing multiple nulls on a large pool of toxicants, checking for coherent characterization across chemically similar compounds.
\textcolor{black}{Note that combination analysis methodology is shared across toxicology and pharmacology.
Due to this overlap, it is often framed in terms of "drugs" rather than "toxicants," "compounds," or "pollutants."
In the remainder, we shall use the term most appropriate to each context.}

Among the most widely adopted baselines \textcolor{black}{for additivity},
the Bliss independence criterion assumes that drugs act through stochastically independent mechanisms \citep{Foucquier2015_landscape}.
{Expressing the response as the probability of an adverse outcome or a fractional effect re-scaled to $[0,1]$, the complementary survival probability under joint exposure thus equals the product of the individual ones.}
Loewe additivity takes the opposite stance \citep{Foucquier2015_landscape}, assuming drugs to have similar mechanisms of action, so that equal-effect doses are mutually interchangeable.
A key limitation is that plain Loewe additivity assumes the relative potency to be constant across effect levels.
Extensions of Loewe additivity (Loewe$^+$) relax this requirement while retaining the conceptual appeal of dose equivalence \citep{Tallarida2016,Lederer2018_EME}.
Alternatively, the Hand model \citep{Sinzger2019_Hand} focuses on instantaneous rates of effect gain, postulating a linear contribution to the local gradient of the dose-response surface.

None of these paradigms is tied to a particular functional form for the single-exposure dose-response curves, although they are sometimes used in combination with parametric formulations.
In the parametric realm, commonly used methods -- such as ZIP \citep{Yadav2015_ZIP}, BRAID \citep{Twarog2016_Braid}, and MuSyC \citep{Meyer2019_MuSyC,Wooten2021_MuSyC} --
embed interaction parameters directly into extended two-dimensional Hill-type curves.
Parametric copula models \citep{Lambert2019_Copula, Hashizume2022_Copula} aim at offering additional flexibility in specifying the joint distribution, though this does not necessarily translate to better performance in practice.
{In the nonparametric realm,
\citet{Ronneberg2021_bayesynergy} and \citet{Shapovalova2022_HandGP} leverage Gaussian processes for a more flexible surface fit, though without enforcing monotonicity and mainly committing to a single null model (Bliss and Hand, respectively).}

Given a null model for additivity, assessing evidence for synergy or antagonism requires a rigorous statistical testing strategy.
Parametric formulations \citep{Yadav2015_ZIP,Twarog2016_Braid,Wooten2021_MuSyC} simplify testing by focusing on dedicated interaction parameters, {possibly extended to a Bayesian setting \citep{Zhang2023_SynBa}}.
Alternatively, common approaches in the literature include evaluating geometric properties of isoboles at fixed effect levels, computing combination indices, or isobole lengths \citep{Rider2018_ChemicalMixtures}.
However, such metric-based approaches often stop short of rigorous statistical testing procedures.
\citet{Lederer2019_Synergy_index} proposed forming integral contrasts between observed and null-model isoboles, or considering parametric perturbation of generalized Loewe additivity via a single synergy index -- both coupled with bootstrap-based inference.
Yet, such single indices rigidly constrain synergy or antagonism to act uniformly across the entire dose space,
while integral-based contrasts quantify deviations on a scale that need not correspond to the one underlying the data-generating process or full likelihood.

\vspace{-5pt}

\subsection{Contributions and Main Findings}

In overcoming these limitations, our contribution to the analysis of the \textit{C. elegans} assay is threefold.
First, we model the full distribution of the ordinal damage score as a convex combination of two extreme profiles, corresponding to null and maximal effect. 
The mixing weight between them is driven by the toxicant doses, so that inference on this single scalar index recovers the entire dose-response surface while fully preserving the ordinal nature of the observed scores. 
Second, we avoid rigid parametric assumptions on both single- and joint-exposure profiles, relying instead on flexible representations built from monotone splines. For the joint exposure, we cast the problem in a radial coordinate system, which we find to reconstruct the dose-response surface more faithfully than alternative flexible approaches based on copulas.
Third, we assess additivity through a formal likelihood ratio test, benchmarking the fitted joint effect against popular null models of additivity as the null hypothesis.
We use a parametric bootstrap to approximate the distribution of the test statistic and thereby rigorously quantify the evidence for synergy or antagonism.

Applied to the \textit{C. elegans} assay, our framework uncovers evidence of synergy between BaP and Pb.
Their joint neurotoxic effect exceeds the additive baselines, strongly so against
{Loewe$^+$ and Hand null models, and}
moderately against the more conservative Bliss criterion. 
This departure from additivity is not uniform across the dose space but is concentrated at moderate BaP combined with high Pb doses -- 
a localized, asymmetric signal that the flexibility of our formulation is key to resolving.

The remainder of the paper is organized as follows. Section~\ref{sec_data} describes the \textit{C. elegans} assay data, the ordinal damage scores, and the targeted exposure design.
Section~\ref{sec_method} details the statistical methodology employed.
Section~\ref{sec_results} analyzes BaP--Pb additivity in the \textit{C. elegans} assay.
Section~\ref{sec_discussion} concludes with a discussion, while further benchmarks are investigated in the Appendices.

\vspace{-5pt}

\section{The \textit{C. elegans} Assay Data}\label{sec_data}

{
Our assay exposes \textit{C. elegans} nematodes to the two toxicants, both individually and jointly, causing structural and functional damage to individual neurons.
We capture neuronal morphology through fluorescence microscopy and quantify neuron-level damage via the ordinal scale of \citet{bijwadia2021quantifying}, illustrated in Figure~\ref{fig_neuron_photos}. 
This scale system ranges from 0 for a neuron of intact morphology to $L=6$ for one whose structural integrity is entirely lost, and it grades severity by combining fluorescence intensity with structural cues such as the number of breaks along the neuron. 
}
The resulting numerical damage scores are strictly ordinal in nature, rather than continuous.
While a higher score indicates greater neuronal damage,
equal spacing between categories cannot be assumed.
%%%%%%%
The assay data consists of $N=7558$ total neuron-level damage
scores $y_i \in \{0,\dots,L\}$, belonging to worms exposed to varying dose concentrations $d_{1i} \!\in\! [0, 10]\, (\mu M)$ of benzo[a]pyrene and $d_{2i} \!\in\! [0, 120] \, (\mu M)$ of lead acetate.
{A complete description of the experimental protocol -- from strain maintenance through microscopy and scoring -- is provided in Appendix~\ref{app_exp_protocol}.}
We refer to the exposure as {\it joint} when both $d_{1i}$ and $d_{2i}$ are non-null, and {\it single} when only one of $d_{1i}$ and $d_{2i}$ is null.
Control measurements correspond to $d_{1i}=d_{2i}=0$.
The specific numbers of neurons probed at each dose combination are reported in Table~\ref{tab_dose_combos_N}.

% \vspace{-10pt}

The assay comprised two main chunks of data: a first set of measurements entailing only single exposures to Pb or BaP, and a second one comprising both single and joint exposures.
The first set of data was used to create a preliminary estimate of the single toxicant's effects.
In turn, these were used to construct a uniform ray design in a dose-aligned space, determining target dose combinations to be probed in the second data chunk.
A visual illustration of such a design is provided in Figure~\ref{fig_data_ray_design}, while more detailed information is provided in Appendix~\ref{app_ray_design}.
From the panels A and C of Figure~\ref{fig_data_ray_design}, it is evident that the damage profile induced by the highest measured dose of lead acetate is considerably less severe than that observed at the maximum dose of BaP.
This is not due to a dose-response saturation effect in the biological sense, but rather to a chemical limitation: above the maximum considered concentration, lead acetate reaches the limit of solubility in the liquid culture medium.
Configurations with higher concentrations of Pb cannot thus be probed in our experimental setup.

\newpage

\begin{figure}[ht!]
    \vspace{-5pt}
    \centering
    \includegraphics[width=0.9\linewidth, trim=30 10 720 30, clip]{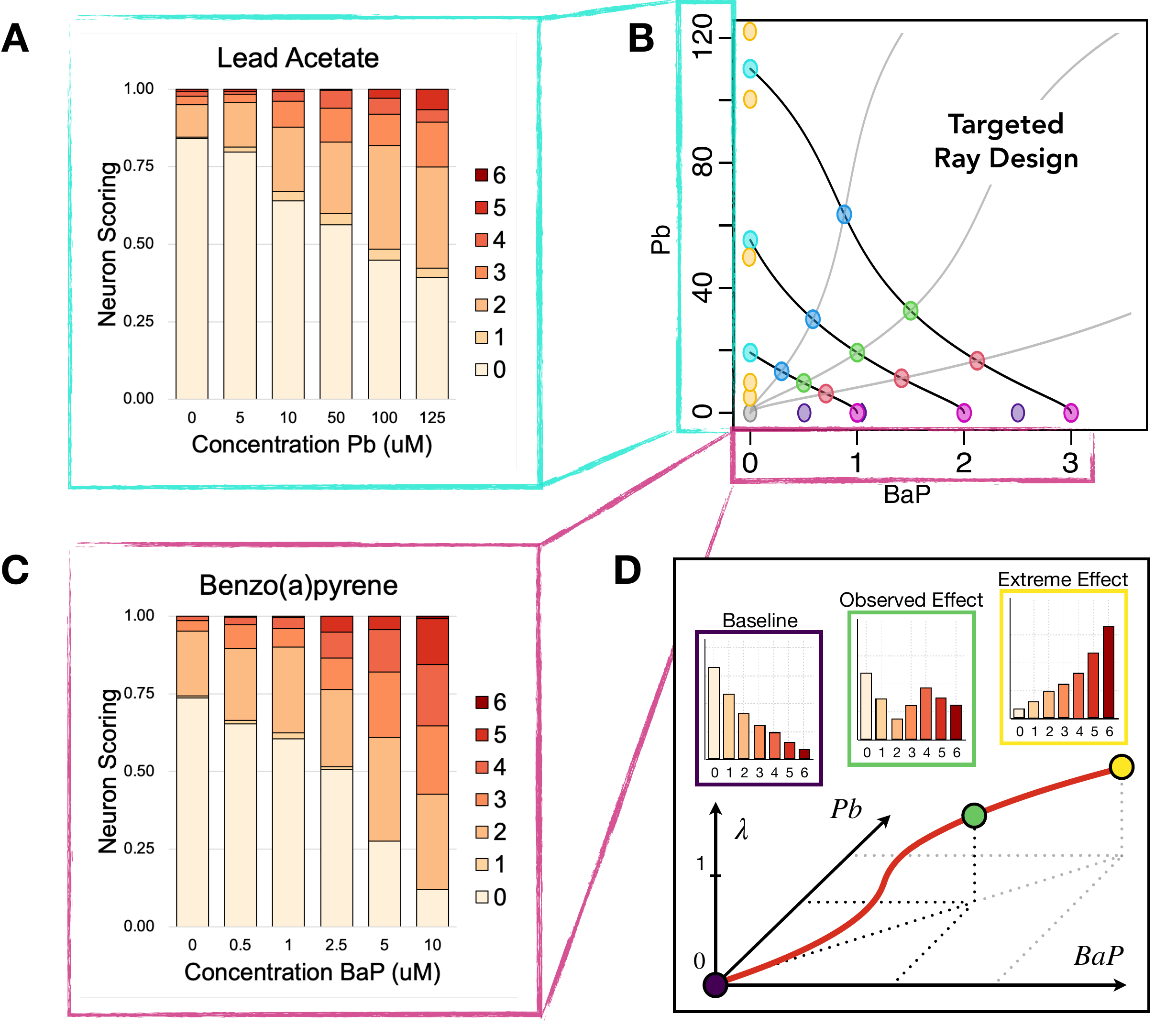}
    \vspace{-10pt}
    \caption{
    Overview of the {\it C. elegans} co-exposure assay and proposed modeling rationale. 
    (A,C) Distribution of neuron-level damage scores, shown as stacked proportions across increasing single-exposure concentrations of lead acetate (A; Pb, 0–125 $\mu$M) and benzo[a]pyrene (C; BaP, 0–10 $\mu$M).
    Both compounds induce a dose-dependent redistribution of mass toward the more severe categories.
    The markedly weaker maximum effect for Pb is due to
    the limit of solubility in the liquid culture medium.
    %%%%%%%%%%%%%%%%%
    (B) Targeted ray design in the (BaP, Pb) dose space. Preliminary single-exposure estimates were used to place joint dose combinations along rays of fixed composition.
    Gray lines trace the fixed-ratio rays (points sharing a color lie on a common ray), and black curves the corresponding iso-effect contours.
    %%%%%%%%%%%%%%%%%
    (D) Schematic of the proposed modeling approach, where a single scalar weight $\lambda \in [0,1]$ quantifies damage severity, determining the full outcome distribution of the ordinal scores.
    $\lambda$ varies smoothly over dose space, interpolating the score distribution from a baseline regime ($\lambda \rightarrow 0$, unexposed; purple) through intermediate observed effects (green) to an extreme-effect regime ($\lambda \rightarrow 1$; yellow).
    }
    \label{fig_data_ray_design}
\end{figure}

\setcounter{savedfigure}{\value{figure}}
\renewcommand{\figurename}{Table}
\addtocounter{table}{0}
\setcounter{figure}{\value{table}}
\begin{figure}[ht!]
\centering
\begin{tabular}{l | cc | cc}
\toprule
& \multicolumn{2}{c|}{1-on-1 Unadjusted Tests} & \multicolumn{2}{c}{Multiple-testing Adjusted} \\
\cmidrule(lr){2-3} \cmidrule(lr){4-5}
& \makecell{Wilcoxon} & \makecell{KS} 
& \makecell{Wilcoxon} & \makecell{KS} \\
\midrule
Raw Data      & 43\% & 43\% & 24\% & 29\% \\
Batch Adjusted & 19\% & 14\% & 0\%  & 0\%  \\
\bottomrule
\end{tabular}
\caption{Percentage of control-group pairs with statistically significant pairwise differences across data rays. 
Each column reports results for a different testing procedure -- Wilcoxon rank-sum (Wilcoxon) and permutation-based Kolmogorov--Smirnov (KS) -- on raw and batch-corrected data, with and without Bonferroni correction across pairs.}
\label{tab_Wilcoxon}
\end{figure}
\renewcommand{\figurename}{Fig.}
\setcounter{figure}{\value{savedfigure}}

In both data chunks, observations are grouped in sub-batches, each one spanning one ray $r_i$ of fixed dose ratio -- corresponding to either joint or single exposures.
Each sub-batch includes control measurements of unexposed worms. 
Control groups allow a direct comparison between batches, not confounded by the dose-response effects thanks to the lack of exposure.
Related studies have highlighted significant statistical differences across independent experimental replicates in these types of toxicological assays \citep{Presman2026_BayesRank}.
Our data also show evidence of such batch effects, quantified via Wilcoxon and Kolmogorov-Smirnov tests comparing control groups across arms.
This is also visually appreciable by comparing the leftmost columns in panels A and C of Figure~\ref{fig_data_ray_design}.
We addressed batch effects correction via a pre-processing routine, which proves effective for our purposes while avoiding modeling complications.
Table~\ref{tab_Wilcoxon} supports the effectiveness of our approach, while full details of the procedure are presented in Appendix~\ref{app_pre_process}. 
Accordingly, we omit the sub-batch index $r_i$ from the notation in most of what follows.

\section{Methods}\label{sec_method}

To retain the full information encoded in the ordinal neuron damage scores
$y_i$, we leverage the convex mixture regression framework of
\citet{Canale2018_CoMiRe}.
This construction smoothly interpolates the full distribution
of $y_i$ between two extreme profiles corresponding to the unexposed and maximally-exposed settings.
Conditioned on exposure to a given dose $d_{s} \in [0,D_s^{(\max)}]$ of chemical $s=\{1,2\}$, the cumulative distribution function (CDF) of the response is modeled as
\begin{equation}
  \mathbb{P}[ y_i \leq \ell \mid d_s=d_{si} ] = F_s(\ell \mid d_{si}) =
  \bigl(1 - \lambda_s(d_{si})\bigr)\,F_o(\ell) +
  \lambda_s(d_{si})\,F_\infty(\ell) \;,
  \label{eq_convex_mixture_1D}
\end{equation}
where $F_o$ and $F_\infty$ are the CDFs at zero and maximal exposure, respectively, while $\lambda_s : [0, D_s^{\scriptstyle{(\max)}}] \to [0,1]$ is a weight function capturing the dose-response relationship.
Intuitively, at zero exposure the response follows $F_o$, reflecting the baseline score profile of unexposed worms.
At maximal exposure $D_s^{(\max)}$ it follows $F_\infty$, capturing the most severe damage.
At intermediate doses, the response follows a middle ground that interpolates smoothly between these two extremes, with $\lambda_s(d_s)$ governing how far along this continuum the distribution has shifted.
Note that $F_\infty$ need not represent the maximum conceivable damage level, but simply a profile no less severe than that observed at the highest probed dose.

This formulation is particularly convenient because it recasts the problem in terms of a scalar dose-response index $\lambda_s(\cdot)$, directly amenable to several additivity frameworks.
In the toxicology literature, $\lambda_s(\cdot)$ is often assumed to follow a Hill-type curve, such as
\begin{equation*}
    \lambda_s(d_s)=\frac{\omega_o+\omega_s \, \big( d_s/c_s \big)^{\,h_s}}{1+ \big( d_s/c_s \big)^{\,h_s}}
    \qquad \text{with} \quad 0 \leq \omega_o \leq \omega_s \leq 1 ,\; c_s>0,\; h_s>0 \;.
\end{equation*}
However, we want to avoid restricting parametric assumptions of this sort to accommodate possibly non-standard dose-response profiles.
Thus, we only require $\lambda_s(\cdot)$ to be a monotone non-decreasing weight function, satisfying $\lambda_s(0) = 0$ and $\lambda_s(D_s^{(\max)}) = 1$.
To balance flexibility and tractability, we parameterize $\lambda_s$ via integrated monotone splines (I-splines) \citep{Ramsay1988_ISplines}.
% the basis expansion  $\lambda_s(d_s) = \sum_{h=1}^{K} \nu^{(s)}_h \phi_h(d_s)$, where $\big\{\phi_h(\cdot) \big\}_{h=1}^{K}$ are monotone spline basis functions.

% \vspace{-10pt}

\begin{figure}[ht!]
    \centering
    \includegraphics[width=\linewidth, trim=50 250 50 350, clip]{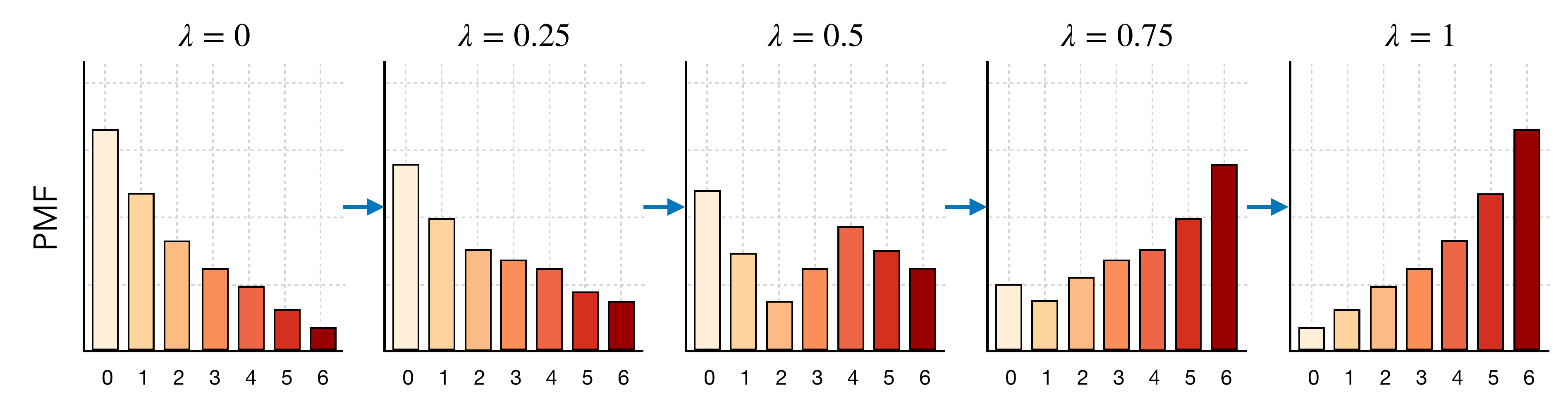}
    \vspace{-25pt}
    \caption{
    Schematic representation of the smooth interpolation of the response distribution. The probability mass functions shift towards high damage score for increasing mixing weight $\lambda \in [0,1]$.
    At $\lambda=0$ the distribution matches the unexposed baseline $F_o$, concentrating mass on low scores; at $\lambda=1$ it matches the maximal-exposure profile $F_\infty$, concentrating mass on high scores. 
     Intermediate values interpolate smoothly between the two extremes, progressively shifting mass toward more severe categories.
    }
    \label{fig_smooth_mix_sketch}
\end{figure}

% \vspace{-10pt}

Recall that chemical $s=1$ stands for BaP and chemical $s=2$ for Pb.
Rather than having two unconstrained weight functions, we additionally set
\begin{equation}\label{eq_l2_g_t2}
    \lambda_2(d_2)= \lambda_1 \big(D_1^{(\max)} \, \tau_2 \, g(d_2) \big) \;,
\end{equation}
without any loss of generality.
Here, the function $g : [0,D_2^{(\max)}] \rightarrow [0,D_1^{(\max)}]$ maps each dose of Pb to the dose of BaP producing an equivalent effect, effectively placing the two chemicals on a common dose scale.
The scalar $\tau_2 \in (0,1)$ accounts for the fact that the maximum observed effect of Pb is less severe than that of BaP, compressing the aligned Pb doses accordingly.
The function $g(\cdot)$ is still constrained to be monotone non-decreasing via a monotone spline basis expansion, which also enforces $g(0) = 0$ and $g(D_2^{(\max)})= 1$.
This alignment has a practical appeal, as it provides a transformed dose space in which the two chemicals are equivalent in terms of their neurotoxic effect -- when administered in isolation.
As detailed in Appendix~\ref{app_ray_design}, we further exploit this alignment to construct the targeted ray design on a principled basis.

\vspace{-5pt}

\subsection{Baselines for Additivity: Weight Function Under Joint Exposure}

\begin{figure}[H]
    \vspace{-5pt}
    \centering
    \includegraphics[width=0.6\linewidth, trim=280 200 250 225, clip]{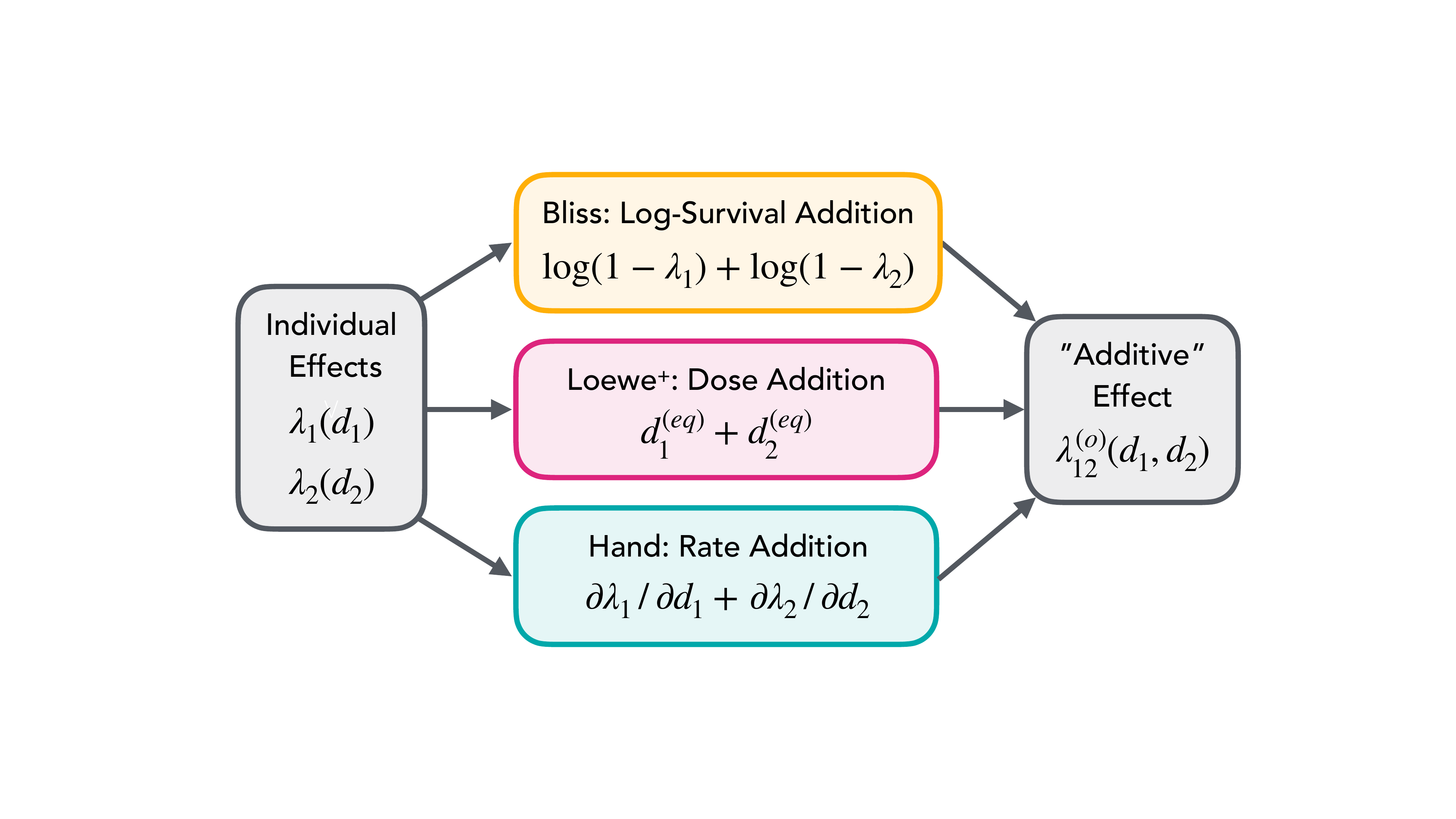}
    \put(-270,130){\makebox(0,0){\textbf{A}}}
    \put(-3,130){\makebox(0,0){\textbf{B}}}
    \includegraphics[width=0.39\linewidth, trim=400 160 640 240, clip]{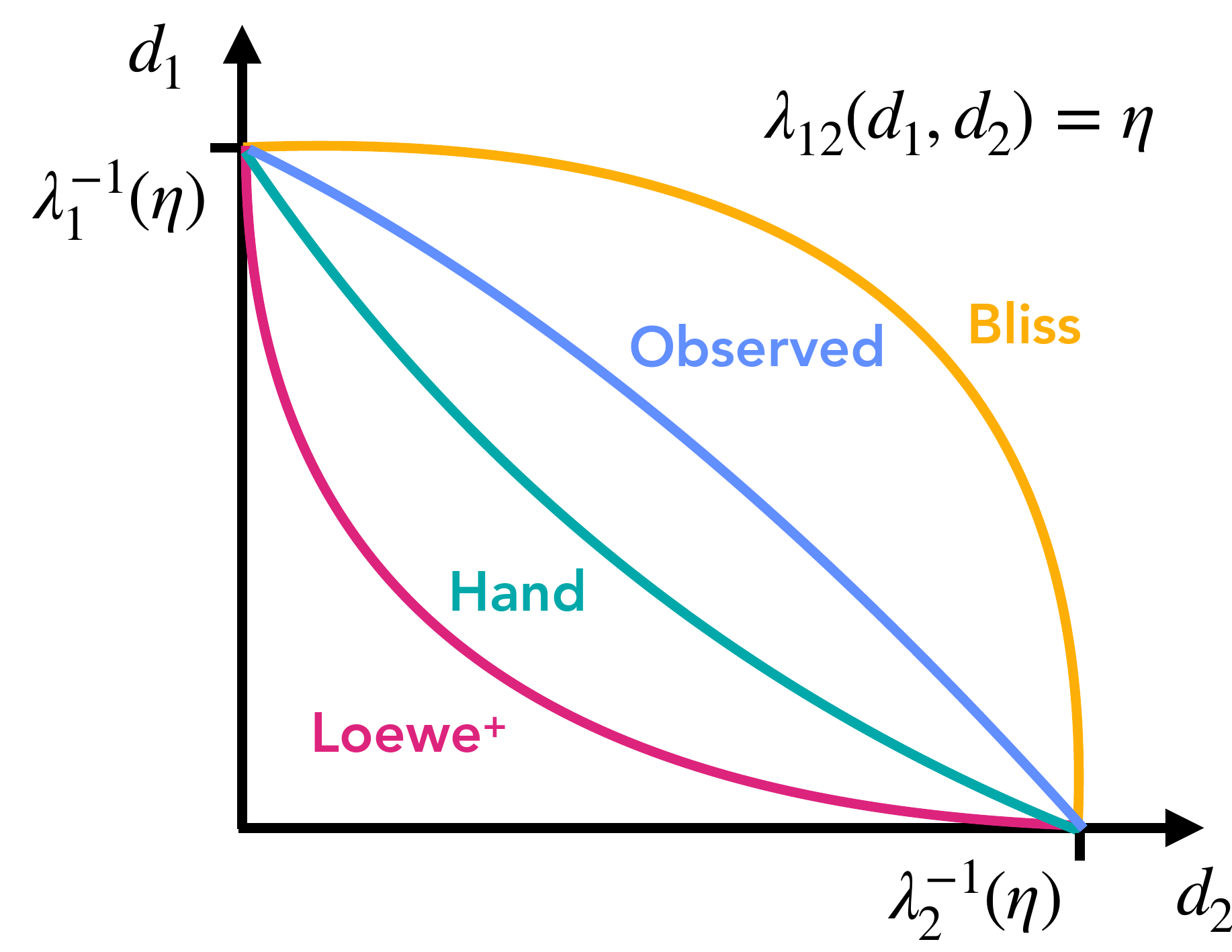}
    \vspace{-5pt}
    \caption{
    Additivity between two toxicants is defined relative to a chosen combination scale, rather than being a universal notion.
    (A) All considered null models combine the individual dose–response weights $\lambda_1(d_1)$ and $\lambda_2(d_2)$ into an additive baseline $\lambda_{12}^{(o)}(d_1,d_2)$, but differ in what quantity is added.
    Bliss's independent action is also referred to as response addition in the toxicological literature. 
    {Viewing $\lambda_s$ as a failure probability (i.e. shift toward more severe damage), this corresponds to adding log-survivals.}
    Loewe$^+$ sums up doses after rescaling them to a common effect scale, while Hand adds instantaneous effect rates along rays.
    %%%%%%%%%
    (B) Deviations from additivity can be visually assessed by comparing isoboles -- curves identifying the dose combinations that produce a common effect level $\eta \in [0,1]$, all sharing the single-exposure anchors $\lambda_1^{-1}(\eta)$ and $\lambda_2^{-1}(\eta)$ on the axes but bowing differently in between.
    An observed isobole lying closer to the origin than a given null isobole indicates synergy relative to that null, since the effect level $\eta$ is reached at lower combined doses. An observed isobole lying further out indicates antagonism, reaching $\eta$ only at higher combined doses. 
    The same observed isobole can therefore be synergistic with respect to one baseline yet antagonistic with respect to another.
    In the sketch above, the observed isobole would be read as synergistic relative to Bliss, yet antagonistic relative to Loewe$^+$ and Hand.
    Equivalently, in a 3D plot of the effect surface over the dose plane, synergy corresponds to the observed surface rising above the null and antagonism to its falling below.
    }
    \label{fig_additivity_schematic}
\end{figure}

The proposed framework readily extends to joint exposure by letting the weight function depend on both chemicals' doses
\begin{equation}
\begin{aligned}
    \mathbb{P}[ y_i \leq \ell \mid d_1=d_{1i}, d_2=d_{2i} ] 
    &= F_{12}(\ell \mid d_{1i},d_{2i}) \\
    &= \big(1-\lambda_{12}(d_{1i},d_{2i})\big) \, F_o(\ell) + \lambda_{12}(d_{1i},d_{2i})\, F_\infty(\ell) \;. 
\end{aligned}
\label{eq_convex_mixture_2D}
\end{equation}
Here, $\lambda_{12} : [0,D_1^{(\max)}] \times [0,D_2^{(\max)}] \rightarrow [0,1]$ is required to be monotone non-decreasing in each argument, and with $\lambda_{12}(0,0)=0$.
Additionally, the joint-exposure weight function must recover the individual dose-response weight functions at the boundaries, so that
\begin{equation*}
    \lambda_{12}(d_1,0)=\lambda_{1}(d_1)
    \qquad \text{and} \qquad
    \lambda_{12}(0,d_2)=\lambda_{2}(d_2).
\end{equation*}
Different approaches to combination analysis are essentially reflected in distinct specifications of $\lambda_{12}(\cdot \,,\cdot)$, constructed by combining the single-exposure weight functions $\lambda_1(\cdot)$ and $\lambda_2(\cdot)$ under different notions of additive baseline behavior.
Figure~\ref{fig_additivity_schematic} illustrates schematically how additivity can be defined at different scales, while below we detail the frameworks considered in our analysis.
This includes the most popular approaches grounded in general and broadly applicable principles, while we discard again those intrinsically tied to specific parametric formulations.
Following recommended practice \citep{DiVeroli2016_Combenefit,Ianevski2017_SynergyFinder}, we shall assess additivity against each of these different null models, providing a comprehensive picture of the empirical evidence.
Henceforth, we will use the superscript ``$(o)$" to highlight $\lambda_{12}^{(o)}(\cdot \,,\cdot)$ being a function of the single-drug exposure weight profiles as prescribed by a given null model for additivity.

\newpage

\begin{itemize}
    %%%%%%%%%%%%%%%%%%%%%%%%%%%%%%%%%%%%%%%%%%
    \item[\textbf{(Bliss)}] Under Bliss independence \citep{Foucquier2015_landscape}, the two toxicants are assumed to act through stochastically independent mechanisms, with no direct interaction between their respective contributions to the response distribution.
    {
    To this end, the weight $\lambda_s(d_s) \in [0,1]$ can be seen as the probability of the ``failure'' event $A_s$ that toxicant $s$ drives neurons toward the more severely damaged regime, with $1-\lambda_s(d_s)$ being the corresponding survival probability.
    Under independence of $A_1$ and $A_2$, the joint survival 
    $1 - \lambda_{12}^{(o)}(d_1,d_2) = (1-\lambda_1(d_1))(1-\lambda_2(d_2))$
    factorizes as the product of the individual survivals.
    Equivalently, the inclusion-exclusion principle 
    $\mathbb{P}(A_1 \cup A_2) = \mathbb{P}(A_1) + \mathbb{P}(A_2) - \mathbb{P}(A_1)\mathbb{P}(A_2)$ gives the multiplicative rule}
    \begin{equation*}
      \lambda_{12}^{(o)}(d_1, d_2) =
      \lambda_1(d_1) + \lambda_2(d_2) -
      \lambda_1(d_1)\,\lambda_2(d_2).
      % \label{eq_bliss}
    \end{equation*}
    %%%%%%%%%%%%%%%%%%%%%%%%%%%%%%%%%%%%%%%%%%
    \item[\textbf{(Loewe$^{+}$)}] Loewe additivity and its extensions assume the two toxicants act through interchangeable mechanisms, so that doses combine additively once rescaled to a common effect scale.
    For any given dose $d_2$ of drug 2, its equivalent dose of drug 1 if defined as $d_1^{\mathrm{eq}}(d_2) = \lambda_1^{-1}(\lambda_2(d_2))$, and represents the dose of drug 1 that produces the same effect as $d_2$ -- with an equivalent definition for $d_2^{\mathrm{eq}}(d_1) = \lambda_2^{-2}(\lambda_1(d_1))$.
    Accordingly, the \emph{Explicit Mean Equation} (EME) of \citet{Lederer2018_EME} constructs the additive null as the average of the effects obtained by augmenting each drug with the equivalent dose of the other
    \begin{equation*}
          \lambda_{12}^{(o)}(d_1, d_2)
          \;=\;
          \tfrac{1}{2}\,\lambda_1\!\Big(d_1 + \lambda_1^{-1}\big(\lambda_2(d_2)\big)\Big)
          +
          \tfrac{1}{2}\,\lambda_2\!\Big(d_2 + \lambda_2^{-1}\big(\lambda_1(d_1)\big)\Big)
          \; .
    \end{equation*}
    This generalizes classical Loewe additivity \citep{Foucquier2015_landscape}, where the \emph{General Isobole Equation} (GIE) characterizes the set of dose combinations $\big\{(d_1,d_2)\big\}_{S(\eta)}$ that produce any given effect level $\eta \in [0,1]$ under additivity as a linear segment $S(\eta)$ in the unaligned dose space
    \begin{equation*}
          \frac{d_1}{\lambda_1^{-1}(\eta)} + \frac{d_2}{\lambda_2^{-1}(\eta)} = 1
          \qquad \forall \, d_1 \in \big[0, \lambda_1^{-1}(\eta)\big] \quad
          \textcolor{gray}{\Big( \text{or} \; \forall \, d_2 \in \big[0, \lambda_2^{-1}(\eta)\big] \Big)} 
          \;.
    \end{equation*}
    A key limitation of the GIE is that it implicitly assumes the relative potency among drugs $\chi(\eta) = \lambda_1^{-1}(\eta) / \lambda_2^{-1}(\eta)$ to be constant across effect levels $\eta$. 
    This is equivalent to requiring the two dose-response profiles to be a simple horizontal dilation of one another on the dose scale -- i.e.\ $\lambda_2(d) = \lambda_1(cd)$ for some constant $c > 0$.
    The EME overcomes this restriction while preserving the sham combination principle, whereby a drug combined with itself should produce no interaction effect. \\[-5pt]
    %%%%%%%%%%%%%%%%%%%%%%%%%%%%%%%%%%%%%%%%%%
    \item[\textbf{(Hand)}] The Hand model \citep{Sinzger2019_Hand} also satisfies the sham combination principle, but operates on instantaneous rates of effect gain rather than on doses directly.
    Specifically, it postulates that both toxicants contribute linearly to the local effect gradient along any fixed-ratio ray -- working in the unaligned space. 
    For a dose pair $(d_1, d_2)$ with mixing ratio
    $\nu = d_1/(d_1+d_2)$ and total dose $\rho = d_1+d_2$, the combined effect
    $\psi_{12,\nu}(\rho) := \lambda_{12}\big(\nu \rho, (1-\nu)\rho \big)$ satisfies the ordinary differential equation (ODE)
    \begin{equation*}
        \psi'_{12,\nu} \big(\psi_{12,\nu}^{-1}(\eta)\big) = 
        \nu\, \lambda_1' \big(\lambda_1^{-1}(\eta)\big) +
        (1-\nu)\, \lambda_2' \big(\lambda_2^{-1}(\eta)\big) \;,
        \label{eq:hand}
    \end{equation*}
    with boundary condition $\psi_{12,\nu}(0)=0$.
    The dose-response curve is then recovered as
    \begin{equation*}
    \lambda_{12}^{(o)}(d_1,d_2) = \psi_{12,\,d_1/(d_1+d_2)}(d_1+d_2) \;,
    \end{equation*}
    and can be easily computed numerically along each ray of fixed $\nu$ by rewriting the ODE in integral form.
    {Leveraging the dose alignment function $g(\cdot)$ further simplifies its evaluation, as detailed in Appendix~\ref{app_hand}.
}
\end{itemize}

\subsection{Assessing Additivity via Likelihood Ratio}

To assess deviation from additivity in a statistically rigorous way, we cast the problem as a likelihood ratio test of the hypothesis
\begin{equation*}
    \begin{cases} 
    \,\mathrm{H}_0: \quad \lambda_{12} \in \mathcal{F}_0 = \big\{ \lambda^{(o)}_{12}
     \big\} \\[5pt]
    \,\mathrm{H}_1: \quad \lambda_{12} \in \mathcal{F}_1 
    \; ,
    \end{cases}
\end{equation*}
where $\lambda^{(o)}_{12}(\cdot \,,\cdot)$ is the weight function induced by a given null model of choice, and $\mathcal{F}_1$ is a flexible family of joint weight functions $\lambda_{12}(\cdot \,,\cdot)$ satisfying the boundary and monotonicity constraints detailed above.
Note that the overall log-likelihood reads
\begin{equation}\label{eq_likelihood}
\begin{aligned}
    \mathcal{L}\big(y\,;\lambda_{12}\big) &= \sum_{i=1}^N \log \mathbb{P}[ y_i = \ell_i \mid d_1=d_{1i}, d_2=d_{2i} ] \\
    &= \sum_{\ell=0}^L \sum_{r=1}^R \sum_{(d_{1} , d_{2} ) 
    \in \mathcal{DC}(r)} {{n}_{d_{1} \, d_{2} \, \ell}^{(r)}} \cdot \log \Big (F_{12}(\ell \mid d_{1},d_{2}) - F_{12}(\ell-1 \mid d_{1},d_{2}) \Big) \;,
\end{aligned}
\end{equation}
where $\mathcal{DC}(r)$ is the list of unique dose-pairs values probed along ray $r$, and we introduced the counts
$n_{d_1 d_2\,\ell}^{(r)} = {\sum_{i=1}^{N}} \mathbbm{1}
\big(y_i = \ell \mid d_{1i}=d_1,\;d_{2i}=d_2,\; r_i=r \big)$.
The null distribution of the test statistics
\begin{equation*}
    T(y) = -2\Big( \mathcal{L}\big(y\,;\lambda^{(o)}_{12}\big) - \text{sup}_{\lambda_{12} \in \mathcal{F}_1} \mathcal{L}\big(y\,;\lambda_{12}\big)\!\Big)
\end{equation*}
is nonstandard due to the mixture structure from equation~\eqref{eq_convex_mixture_2D}.
We therefore estimate the distribution of $T(y)$ via parametric bootstrap, generating replicate datasets under either hypothesis without imposing any distributional assumption on the test statistic \citep{McLachlan1987_Bootstrap}.

Crucially, $\mathcal{F}_1$ must be flexible enough to achieve good fit across a wide variety of observable dose-response surfaces, while encompassing the selected $\lambda^{(o)}_{12}$ as a special case -- or at least being able to approximate it with negligible error.
In the literature, an appealing candidate approach is to leverage a bivariate copula $\mathcal{C} : [0,1]\times[0,1] \to [0,1]$ \citep{Nelsen2006_Copulas} to encode the dependence structure between the effects of the two chemicals
\begin{equation}
    \begin{aligned}
        \lambda_{12}(d_1, d_2) 
        &= \lambda_{1}(d_1) + \lambda_{2}(d_2) - \mathcal{C}\big(\lambda_{1}(d_1),\lambda_{2}(d_2)\big) \\
        &= 1- \mathcal{C}\big(1-\lambda_{1}(d_1),1-\lambda_{2}(d_2)\big) \;.
    \end{aligned}
    \label{eq_lambda_copula}
\end{equation}
Here, $\mathcal{C}$ is a bivariate distribution function with uniform marginals, satisfying $\mathcal{C}(u,0) = \mathcal{C}(0,v) = 0$ and $\mathcal{C}(u,1) = u$, $\mathcal{C}(1,v) = v$ for all $u,v \in [0,1]$.
This automatically guarantees that $\lambda_{12}(\cdot \, , \cdot)$ satisfies the monotonicity constraints, takes values in $[0,1]$, and recovers the individual dose-response profiles at the boundary.
Note that Bliss independence arises as the special case of the independence copula $\mathcal{C}(u,v) = uv$, coherently with the intended lack of toxicant interaction.

\textcolor{black}{
Parametric copula families have been proposed in drug-combination analysis
\citep{Lambert2019_Copula, Hashizume2022_Copula}, but are limited by their specific functional form.
A flexible spline-based representation of $\mathcal{C}$ could in principle accommodate a wider range of dependence structures, yet our empirical results in Appendix~\ref{app_Copulas} show that even this fails to reconstruct relevant null baselines $\lambda_{12}^{(o)}$ -- except trivially for Bliss.
This failure may be rooted structural limitation: a valid copula requires non-negative density everywhere -- i.e. $\partial^2 \mathcal{C}/\partial u\, \partial v \geq 0$) -- which a stronger condition than the coordinatewise monotonicity actually needed for $\lambda_{12}$ -- i.e. $\partial {\lambda}_{12}/\partial d_1 \geq 0$ and $\partial {\lambda}_{12}/\partial d_2 \geq 0$.
Quasi-copulas \citep{Stopar2024_QuasiCopulas,Dolinar2024_QuasiCopulas} relax precisely this condition, but lack the constructive characterizations needed for practical statistical modeling.
}

\subsection{Alternative Hypothesis via Unconstrained Radial Fit}\label{sec_radial_splines}

To overcome this limitation, we instead propose to work directly with a spline-based representation of $\lambda_{12}$ in a radial coordinate system within an aligned dose space.
Specifically, we first map the original dose pair $(d_1, d_2)$ to aligned coordinates $(x_1,x_2)$ 
-- placing both chemicals on a common scale through the function $g(\cdot)$ --
and then transform to radial coordinates $(\rho,\nu)$, where
\begin{equation*}
\begin{cases}
x_1 = d_1 \\[3pt]
x_2 = D_1^{(\max)} \, \tau_2 \, g(d_2),
\end{cases}
\qquad \text{and} \qquad \;
\begin{cases}
\rho = x_1 + x_2 \\[3pt]
\nu = x_2 \,/\,(x_1+x_2) \;.
\end{cases}
\end{equation*}
Here, $\rho$ represents the total aligned dose and $\nu \in [0,1]$ the mixing proportion in the aligned space, identifying a given ray -- in contrast to the Hand model, where the analogous quantities are defined in the unaligned space.
Analogously to 
equation~\eqref{eq_l2_g_t2},
we then model
\begin{equation}
    \lambda_{12}(d_1,d_2) = \lambda_1\!\Big(D_1^{(\max)} \, \tau_{12} \, \phi(\rho,\nu)\Big),
\end{equation}
where the effective dose index $\phi(\rho,\nu)$ is specified as
\begin{equation*}
    \phi(\rho,\nu) = \sum_{h=1}^{K_\phi}
    \frac{e^{w_h(\nu)}}{\sum_{h'} e^{w_{h'}(\nu)}}\,\phi_h(\rho),
    \qquad \qquad
    w_h(\nu) = \beta_{h0} + \sum_{m=1}^{K_w} \beta_{hm}\,\delta_m(\nu).
\end{equation*}
Here, $\big\{ \phi_h: [0,R^{(max)}] \rightarrow [0,1]  \big\}_{h=1}^{K_\phi}$ are I-spline basis functions,
ensuring monotonicity of $\phi$ in $\rho$ for any fixed $\nu$, and $\big\{ \delta_m : [0,1] \rightarrow \Re \big\}_{m=1}^{K_w}$ are B-spline basis functions, governing the smooth variation of the mixing weights across rays. 

Intuitively, this recasts the joint effect as equivalent to a single exposure to the reference toxicant (chemical 1) at an effective dose $D_1^{(\max)}\,\tau_{12} \,\phi(\rho,\nu)$, and mediated through its single-exposure weight function $\lambda_1(\cdot)$.
Recall that BaP is chosen here as a reference simply because of its higher measurable effect.
The scalar $\tau_{12} \in (0,1]$ plays an analogous role to $\tau_2$ in the single-exposure alignment: it accounts for the possibility that the maximum effect attainable under joint exposure does not exceed that of BaP alone, compressing the effective dose range accordingly.
This is coherent with our experimental setup, where the probed dose combinations are expected to produce effects closer to the maximum observed for Pb rather than BaP alone.
Extending the analysis to higher effect levels is technically feasible but would require extrapolating beyond the observed Pb dose range, where the null models may be less reliable.

By construction, $\phi(\rho,\nu)$ is monotone non-decreasing in $\rho$
for any fixed $\nu$, guaranteeing monotonicity of $\lambda_{12}$ along
each ray.
Smoothness across rays is governed by the B-spline weights
$w_h(\nu)$, but monotonicity in $d_1$ and $d_2$ is not automatically enforced, and may be violated when moving across rays.
We address this by complementing the log-likelihood with a roughness penalty on negative partial derivatives of $\lambda_{12}$
\begin{equation*}
    \mathcal{P} = \gamma_1 \,
    \Big\| \Big(\frac{\partial \lambda_{12}}{\partial d_1}\Big)^{-} \Big\|_\infty 
    + \gamma_2 \,
    \Big\| \Big(\frac{\partial \lambda_{12}}{\partial d_2}\Big)^{-} \Big\|_\infty \;,
\end{equation*}
where $(x)^{-} = \max(-x, 0)$ denotes the negative part, and
$\|\cdot\|_\infty$ is the $L^\infty$ norm, and $\gamma_1, \, \gamma_2>0 $ regulate the strength of the penalty
\textcolor{black}{Alternatively, a pool adjacent violators algorithm could be applied as a post-processing step on top of an unconstrained fit.}
Exact recovery of the single-exposure profiles would require
$\phi(\rho, 0) = d_1 / D_1^{(\max)}$ and
$\phi(\rho, 1) = \tau_2 \, g(d_2) / D_1^{(\max)}$,
i.e.\ the boundary rays reducing to the aligned single-chemical doses.
We relax this strict requirement, finding that jointly fitting the model to both single-exposure and combination data retains good agreement with the single-exposure profiles, even without such hard constraints.

\section{Testing BaP-Pb Additivity in the \textit{C. elegans} Assay Data}\label{sec_results}

Equipped with the framework developed in Section~\ref{sec_method}, we can formally assess the joint neurotoxic effect of BaP and Pb in the \textit{C.~elegans} assay.
{
The code to fit the models presented above and reproduce the main results and figures of our analysis is openly available at \url{https://github.com/niccoloanceschi/BaP-Pb-additivity}.
}

\begin{figure}[H]
    \centering
    \includegraphics[width=\linewidth, trim=0 200 0 300, clip]{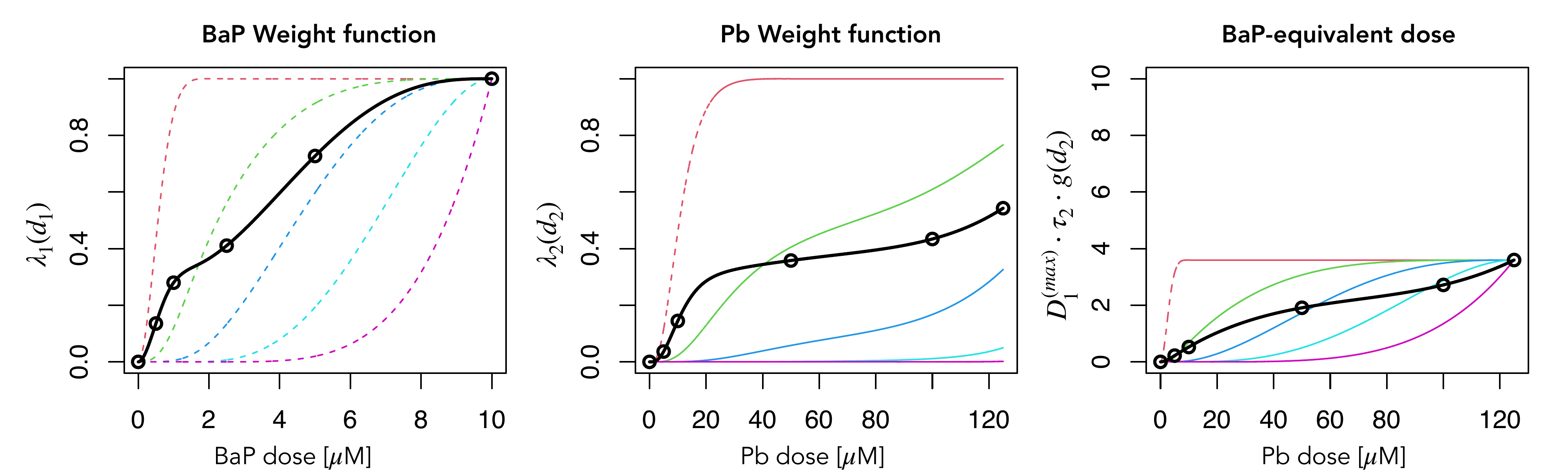}
    \put(-435,135){\makebox(0,0){\textbf{A}}}
    \put(-280,135){\makebox(0,0){\textbf{B}}}
    \put(-130,135){\makebox(0,0){\textbf{C}}}
    \vspace{-20pt}
    \caption{
    Fitted single-exposure weight functions and dose alignment.
    (A–B) Estimated weight functions $\lambda_1(\cdot)$ for BaP (A) and $\lambda_2(\cdot)$ for Pb (B), shown as solid black curves with fitted dose points marked; colored dashed lines are the $K = 5$ I-spline basis elements underlying each parameterization.
    Both profiles are non-monotone in shape and depart from standard Hill-type curves, motivating the flexible spline representation. 
    (C) BaP-equivalent dose mapping $D_1^{(\max)} \, \tau_2 \, g(d_2)$, translating Pb doses onto the aligned BaP scale.
    }
    \label{fig_1D_fit}
\end{figure}

Figure~\ref{fig_1D_fit} reports the fitted single-exposure weight functions $\lambda_1(\cdot)$ and $\lambda_2(\cdot)$, together with the I-spline basis used for their parameterization.
For each spline basis, we used $K=5$ basis elements with fixed knots, leaving the basis expansion weights as the only free parameters.
The estimated shapes exhibit non-standard dose-response profiles that would be poorly captured by parametric Hill-type curves, supporting the use of our flexible spline-based representation.
For the extremal profiles $F_o$ and $F_\infty$ entering the convex mixture in~\eqref{eq_convex_mixture_1D}, we use the most extreme observed empirical CDFs.
Specifically, $F_o$ corresponds to the least severe among all control groups, while $F_\infty$ is taken from the BaP = 10 $\mu$M arm.
While this choice is motivated by modeling simplicity, more
elaborate estimation strategies for $F_o$ and $F_\infty$ could naturally be considered as well.
The estimated $\tau_2$ places Pb's maximal effect at 43\% of the aligned BaP dose range, corresponding to only 54\% of BaP's maximal response weight.
Figure~\ref{fig_1D_fit} also displays the estimated dose-alignment function $g(\cdot)$, mapping each Pb-dose into its BaP-equivalent.
We found the results to be robust to the spline hyperparameters.

% \vspace{-10pt}

% \vspace{-10pt}

\begin{figure}[ht!]
    \centering
    \includegraphics[width=0.9\linewidth, trim=10 50 900 20, clip]{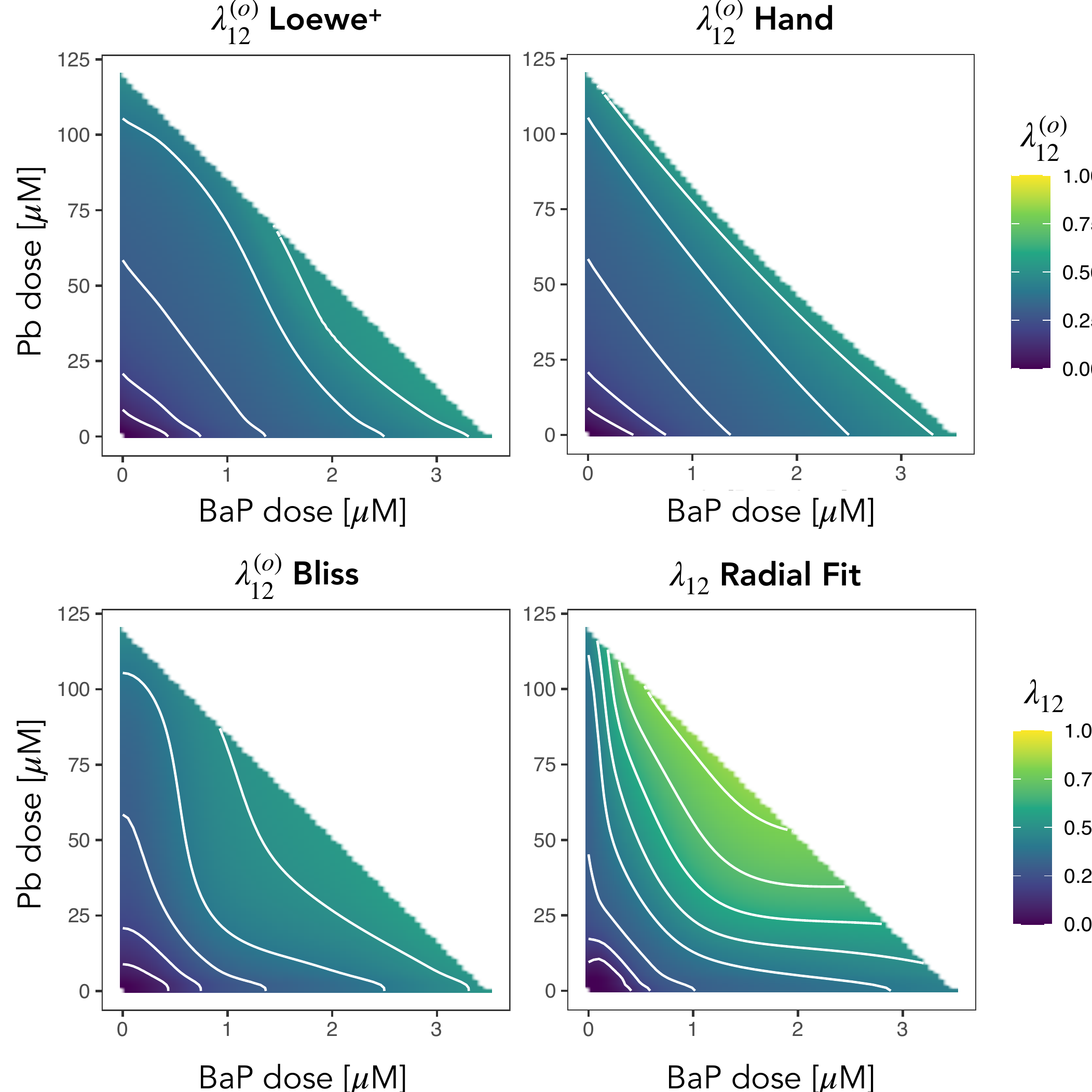}
    \put(-380,415){\makebox(0,0){\textbf{A}}}
    \put(-200,415){\makebox(0,0){\textbf{B}}}
    \put(-380,200){\makebox(0,0){\textbf{C}}}
    \put(-200,200){\makebox(0,0){\textbf{D}}}
    \vspace{-5pt}
    \caption{
    Null models for additivity versus the fitted joint weight surface. 
    {(A–C) Baselines $\lambda_{12}^{(o)}$ for additive joint effect under the Loewe$^+$ (A), Hand (B), and Bliss independence (C) null models -- constructed from the single-exposure fits.}
    (D) Radial spline fit $\lambda_{12}$. 
    In all surface panels, white contours mark the isoboles at weight levels $0.1, 0.2, \dots, 0.9$, tracing dose combinations of equivalent effect.
    Under additivity, the fitted isoboles track those of the null baselines, whereas fitted isoboles lying above (below) the null ones signal synergy (antagonism). 
    In our case, the fitted surface deviates upward from 
    {all}
    baselines at small-to-moderate BaP combined with moderate-to-high Pb (light-green region in D), indicating localized synergy, while the high-BaP/low-Pb region shows weaker evidence -- pointing to an asymmetric synergistic pattern.
    }
    \label{fig_2D_fit}
\end{figure}

\begin{figure}[ht!]
    \centering
    ~\hspace{15pt}
    {\renewcommand{\arraystretch}{1.4}
    \setlength{\tabcolsep}{10pt}
    \begin{tabular}{c||c|c|c}
        \toprule
        $\mathbf{H_0}$ &
        \textbf{Loewe$^+\!$} \textcolor{purple!50}{\scalebox{1.1}{$\blacksquare$}} &
        \textbf{Hand} \textcolor{teal!50}{\scalebox{1.1}{$\blacksquare$}} &
        \textbf{Bliss} \textcolor{amber!50}{\scalebox{1.1}{$\blacksquare$}} \\
        % \midrule
        \Xhline{1pt}
        \textbf{$p$-value} & 0.0002 & 0.0006 & 0.0878 \\
        % \midrule
        \Xhline{0.5pt}
        \textbf{power} & 1.000 & 1.000 & 0.959 \\
        \bottomrule
    \end{tabular} }
    \vspace{10pt}
    \includegraphics[width=\linewidth, trim=40 150 50 0, clip]{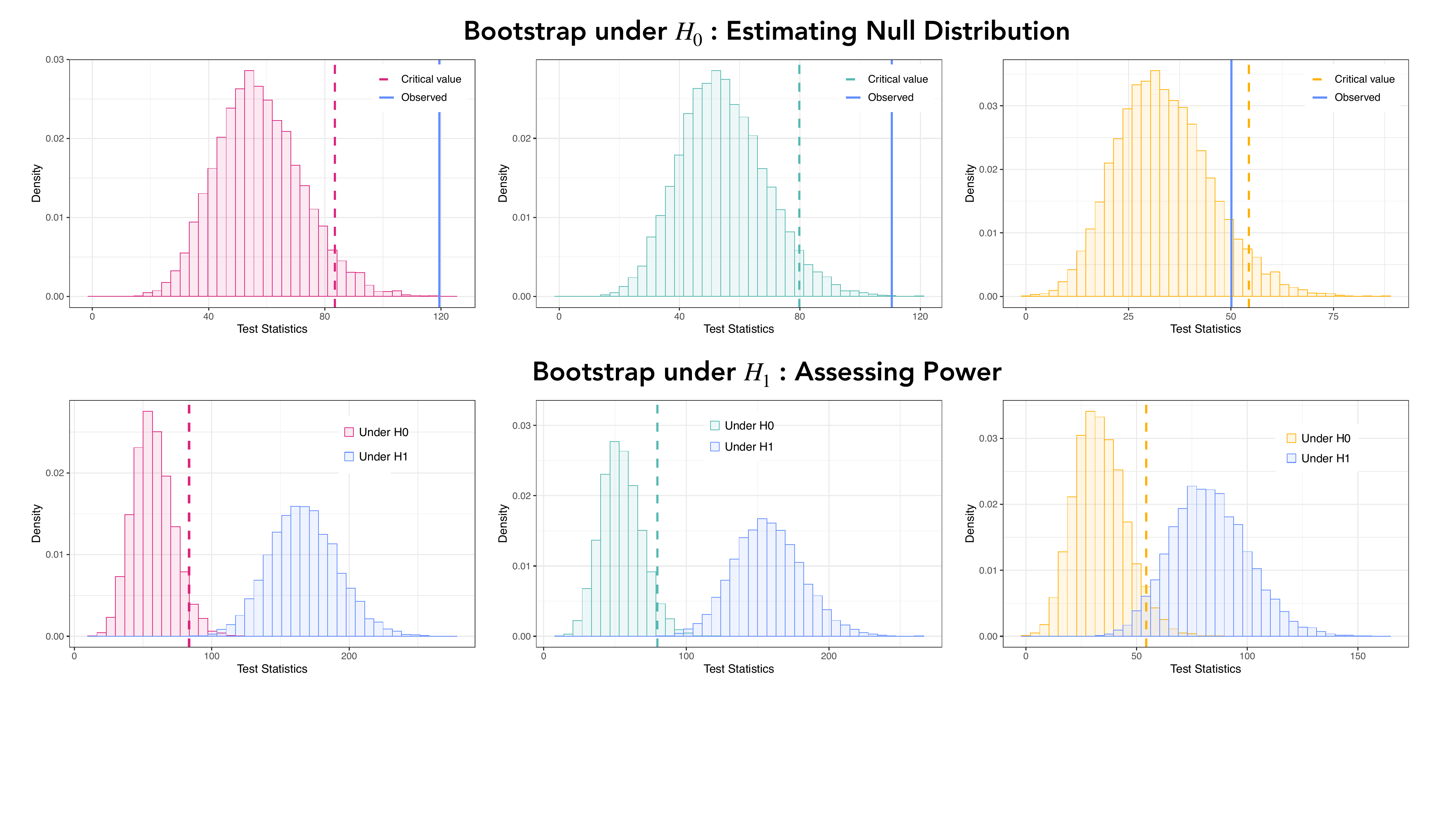}
    \put(-446,205){\makebox(0,0){\textbf{A}}}    
    \put(-288,205){\makebox(0,0){\textbf{B}}}
    \put(-130,205){\makebox(0,0){\textbf{C}}}
    \put(-446,90){\makebox(0,0){\textbf{D}}}  
    \put(-288,90){\makebox(0,0){\textbf{E}}}
    \put(-130,90){\makebox(0,0){\textbf{F}}}
    \vspace{-30pt}
    \caption{
    Likelihood ratio tests of additivity against the Loewe$^+$, {Hand}, and Bliss null models. 
    {(A-C) Null distribution of the test statistic under $H_0$ against Loewe$^+$ (A), Hand (B), and Bliss (C)},
    estimated from 10,000 parametric bootstrap samples; the dashed line marks the critical value at the 0.05 level and the solid blue line the observed statistic $T(y)$. 
    The observed value falls far into the upper tail against Loewe$^+$ {and Hand}, while against Bliss it falls just short of the critical value, corresponding to a p-value marginally above 0.05.
    {(D-F) Distributions of the test statistic simulated under $H_0$ and $H_1$ for Loewe$^+$ (D), Hand (E), and Bliss (F).}
    The weaker evidence against Bliss is consistent with its known conservatism for moderately efficacious combinations.
    }
    \label{fig_bootstrap}
\end{figure}

Given the estimated $\lambda_1$ and $\lambda_2$, we construct the additive baselines $\lambda_{12}^{(o)}$ under Loewe$^+$ and Bliss independence -- defining $H_0$ in our hypothesis test -- and fit the unconstrained radial spline model -- obtaining $\lambda_{12}$ under $H_1$.
Figure~\ref{fig_2D_fit} reports the resulting dose-response surfaces, with white contour lines marking the isoboles,  
identifying dose combinations expected to produce equivalent effects.
Under additivity, the isoboles of the fitted surface $\lambda_{12}$ should closely track those of the chosen null baseline $\lambda_{12}^{(o)}$; a fitted surface lying systematically above (below) the null isoboles signals synergy (antagonism).
Visual inspection reveals a marked upward deviation of the fitted surface from
{all}
null baselines in the region of small-to-moderate BaP doses combined with moderate-to-high Pb doses, corresponding to the light-green patch in panel D of Figure~\ref{fig_2D_fit}.
The complementary region of high BaP and low Pb doses offers less clear visual evidence, suggesting an asymmetric synergistic pattern across the dose space.
{Appendix~\ref{sec_MuSyC} contrasts these results with a popular parametric benchmark (MuSyC), which proves too rigid to capture the localized synergy.}

To formally assess these visual findings, we compare the observed test statistic $T(y)$ against its null distribution estimated via $10{,}000$ parametric {conditional} bootstrap samples generated under each $H_0$ -- whereas simulating under $H_1$ allows us to estimate the power of the test.
The results are summarized in Figure~\ref{fig_bootstrap} and the accompanying tables.
{We find strong evidence of synergy relative to Loewe$^+$ baseline ($p = 0.0002$, power $= 1.000$) and Hand ($p = 0.0006$, power $= 1.000$) baselines.}
Against Bliss independence, we also observe evidence of more-than-additive behavior, though less pronounced ($p = 0.0878$, power $= 0.959$), falling just short of the $0.05$ significance threshold.
The weaker significance against Bliss is consistent with findings in the literature \citep{Wooten2021_MuSyC}, which reported it to be biased toward antagonism in combinations of moderately efficacious chemicals, making synergy harder to detect.

\begin{figure}[ht!]
    \centering
    \includegraphics[width=\linewidth, trim=170 260 120 280, clip]{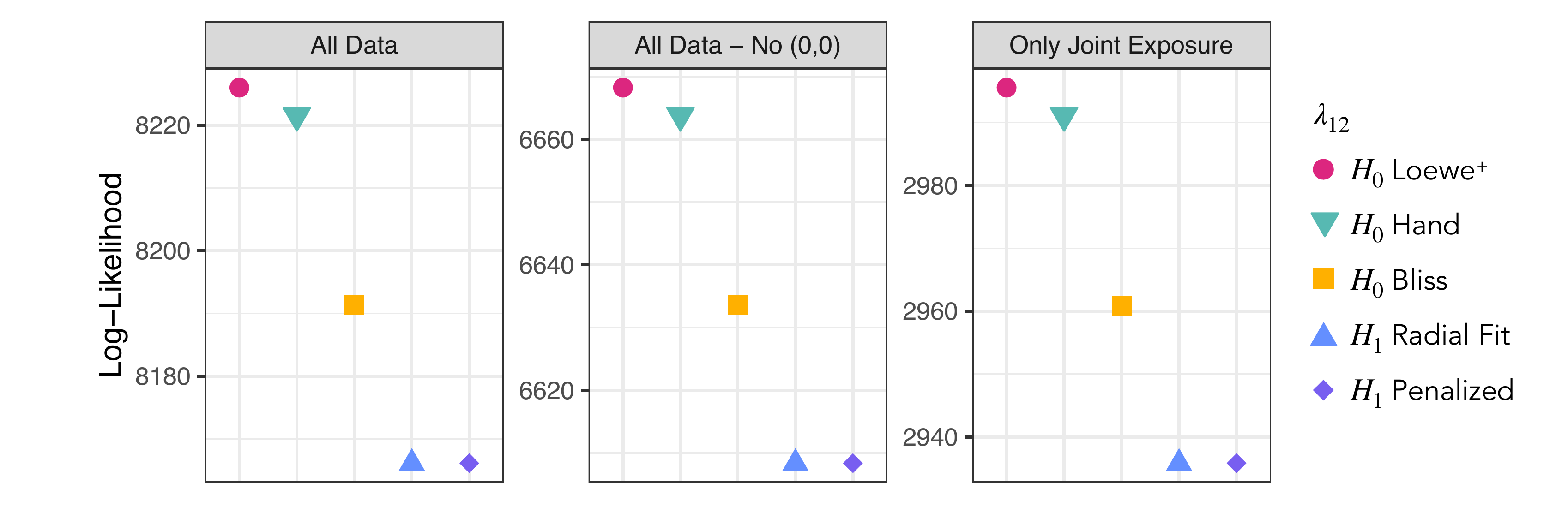}
    \put(-420,140){\makebox(0,0){\textbf{A}}}
    \put(-305,140){\makebox(0,0){\textbf{B}}}
    \put(-190,140){\makebox(0,0){\textbf{C}}}
    \\
    \includegraphics[width=\linewidth, trim=130 190 120 190, clip]{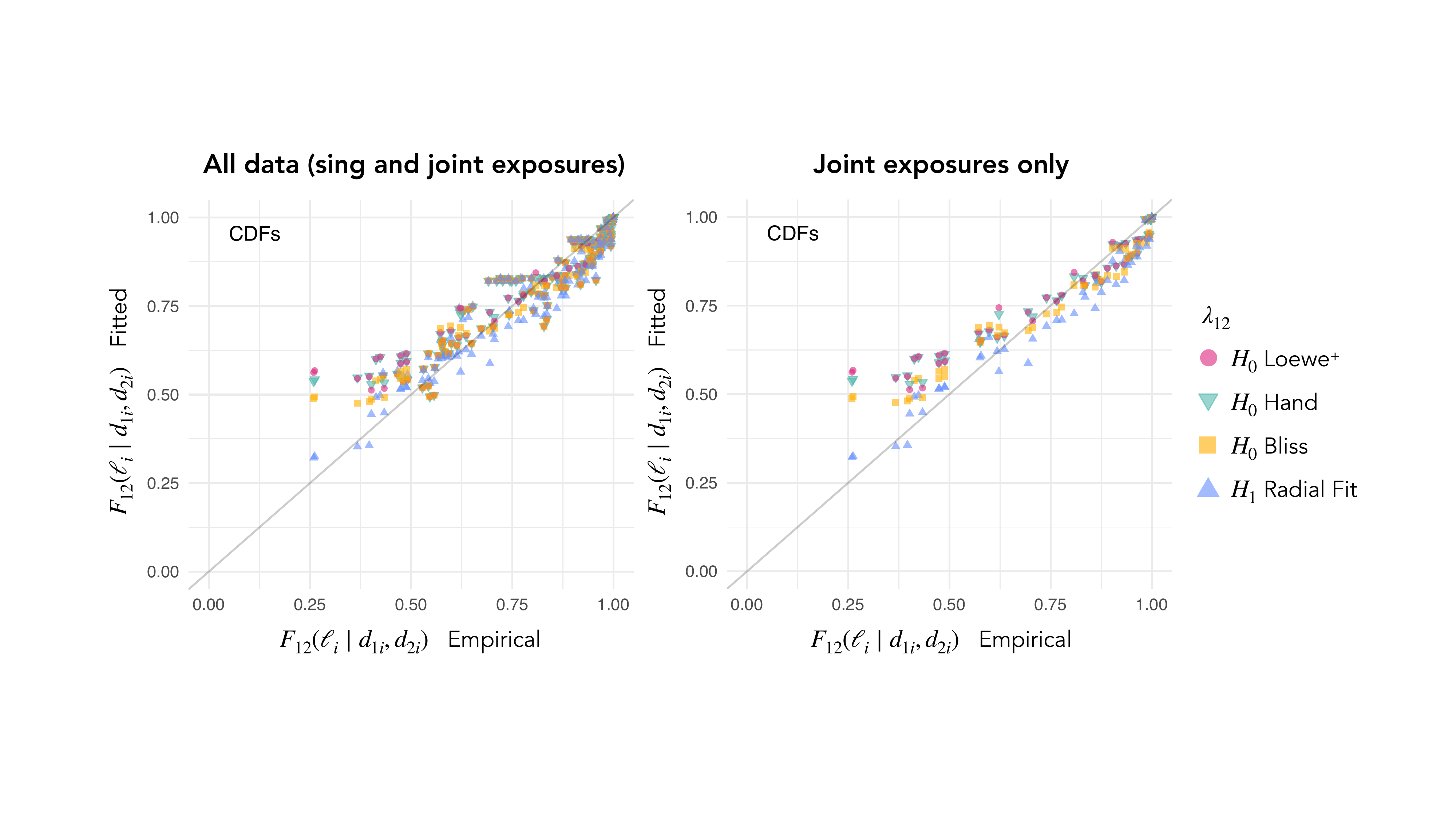}
    \put(-450,190){\makebox(0,0){\textbf{D}}}
    \put(-250,190){\makebox(0,0){\textbf{E}}}
    \put(-395,185){\colorbox{white}{\makebox(120,10){}}}
    \vspace{-20pt}
    \caption{
    Model fit comparisons. 
    (A–C) Log-likelihood attained by each model across data subsets: all data (A), all data excluding controls (B), and joint exposures only (C). 
    The gain from the null models ($H_0$: Loewe$^+$, {Hand}, Bliss) to the radial spline fit ($H_1$) is concentrated in the joint-exposure data.
    The penalized and unpenalized $H_1$ fits nearly coincide, showing that enforcing monotonicity via penalty has little effect on goodness of fit.
    (D–E) Fitted versus empirical CDF values at each observed dose combination, for all exposures (D) and joint exposures only (E). 
    All models roughly scatter around the diagonal, indicating overall adequate fit. 
    The radial fit agrees markedly better with the empirical CDFs for the subset of joint-exposure combinations where the null models depart systematically. 
    These correspond to the lower tails of the ordinal score distributions at higher doses along the blue and green rays in panel C of Figure 5.
    }
    \label{fig_likelihood}
\end{figure}

Figure~\ref{fig_likelihood} provides further insight into the sources of the departure from additivity.
The top panels report the log-likelihood values across models and data subsets, showing that the gain in fit from $H_0$ to $H_1$ is concentrated in the joint-exposure data, with negligible differences on single-exposure observations.
This means that relaxing exact boundary recovery of $\lambda_1$ and $\lambda_2$ in the radial spline fit does not compromise the marginal dose-response profiles.
Moreover, the log-likelihood is barely affected by the
monotonicity penalty, indicating that the penalized and unpenalized fits are in close agreement.

The bottom panels of Figure~\ref{fig_likelihood} display fitted versus empirical CDF values at each observed dose combination -- with seven points per colored dot in the ray design from panel B of Figure~\ref{fig_data_ray_design}, corresponding to the cumulative probabilities evaluated at each of the $L+1 = 7$ response levels.
All models roughly spread around the main diagonal, indicating overall satisfactory goodness-of-fit.
However, the radial fit achieves substantially tighter agreement with the empirical CDFs for a subset of joint-exposure dose combinations where the null models show systematic departures.
Tracing these residuals back to the dose space reveals that they correspond to the lower tails of the outcome distributions for higher dosages along the blue and green rays in panel B of Figure~\ref{fig_data_ray_design} -- or equivalently, to the light-green patch in panel D.
This confirms that the core deviation from additivity originates from moderate BaP combined with high Pb doses.
Gathering more targeted data in this region could sharpen the reported evidence of synergistic signal even further, making it a worthy target for future experimental work.

\section{Discussion}\label{sec_discussion}

In this paper, we studied the joint neurotoxicity of benzo[a]pyrene and lead acetate in a \textit{C. elegans} assay, testing whether their combined action departs from additivity toward synergy or antagonism.
The analysis had to contend with the ordinal nature of the neuron-level damage scores and with dose-response profiles that deviate from standard parametric shapes.
To this end, we modeled the full distribution of the ordinal response as a convex mixture between an unexposed and a maximally exposed profile, with a single dose-driven weight function recasting the problem in terms amenable to established additivity frameworks.
We represented both single- and joint-exposure profiles through flexible monotone splines -- adopting a radial coordinate system for the joint effect -- and assessed additivity through a likelihood ratio test against Loewe$^+$,
{Hand}, and Bliss null models, calibrated via parametric bootstrap.

Applied to the \textit{C. elegans} assay, our analysis 
detects
synergy between the two toxicants: their joint effect exceeds the additive baseline, significantly against Loewe$^+$
{and Hand null models,} 
and more weakly against the Bliss criterion -- which has been reported to bias results towards antagonism \cite{Wooten2021_MuSyC}.
{Our results call into question current regulatory practices, where thresholds are derived compound by compound and joint exposures are evaluated under dose-additivity assumptions.}
The detected signal is asymmetric across the dose space, localized at moderate BaP combined with high Pb doses -- a region that our flexible formulation resolves and that would reward more targeted experimental sampling.

Our framework offers natural extensions in several directions. 
The convex-mixture construction accommodates any ordinal endpoint.
For {\it C. elegans} assays, this includes alternative damage scales, based on non-functional neuron counts \citep{berkowitz2008application, tucci2011modeling}, break-per-neuron counts \citep{gonzalez2014exposure}, damage proportions \citep{hartman2019genetic}, or continuous fluorescence area \citep{luo2019age}. 
More elaborate estimation of the extremal profiles $F_o$ and $F_\infty$ could improve goodness-of-fit even further.
In parallel, \citet{Presman2026_BayesRank} showed that accounting for within-worm dependence among neurons can sharpen inference.
Their hierarchical random effects structure does not carry over directly to our convex mixture formulation, but accommodating such dependence remains a valuable extension to pursue.

{Beyond the specific pair of toxicants considered here, our approach applies to any combination study where a graded response is measured across a dose grid.
Systematic application to other co-occurring exposures could carry direct impact for public health and regulatory practice.
The extension to mixtures of more than two compounds is a priority in this respect, and our formulation is well positioned for this task.
The radial spline representation readily generalizes by enlarging the aligned coordinate system to accommodate additional mixing directions.
This raises a companion question of experimental design, as with three or more chemicals the space of admissible combinations grows far faster than the feasible number of measurements.
Optimal design strategies become particularly crucial for higher-dimensional mixtures, and expanding on the two-compound literature discussed in Appendix~\ref{app_ray_design} is an important direction for future work.
}
{
Finally, our framework could be integrated with a complementary line of research \citep{Wang2021_tensor,Ronneberg2023_GP_matrix_completion,Huusari2025_kernel_based,Kuru2026_DeepNN} that shifts the goal from analyzing a single experiment to predicting dose-response surfaces for untested compound pairs from chemical and cell-line features, trained on a large corpus of previously assayed combinations.
}

\section*{Acknowledgments}
This project has received funding from the United States National Institutes of Health (R01Al167850, R35ES035049, P42ES010356, and R01ES035625).
{The authors thank Michael Aschner for providing the BY200 \textit{C. elegans} strain.}

\putbib
\end{bibunit}

\begin{bibunit}

%%%%%%%%%%%%%%%

\newpage
\setcounter{page}{1}

~\\

\begin{center}
{\sffamily\bfseries\Large{Supplementary Materials for \\[3pt]
``Testing Additivity in Lead and Benzo[a]pyrene--induced\\[3pt]
Neurodegeneration in \textit{Caenorhabditis elegans}''}}\\
\vspace{10pt}
{\sffamily Niccol\`o Anceschi, Javier Huayta, Joel N. Meyer,
David B. Dunson, and Amy H. Herring}
\end{center}

\appendix
\renewcommand{\thesection}{\Alph{section}}
\renewcommand{\thesubsection}{\Alph{section}.\arabic{subsection}}
\renewcommand{\theequation}{\Alph{section}.\arabic{equation}}
\renewcommand{\thefigure}{\Alph{section}.\arabic{figure}}
\setcounter{section}{0}
\setcounter{subsection}{0}
\setcounter{equation}{0}
\setcounter{figure}{0}

{\section{Experimental Protocol for the \textit{C. elegans} Neurotoxicity Assay}\label{app_exp_protocol}}

This section describes the experimental protocol used to generate the analyzed data.

% \vspace{-5pt}

\paragraph{{\it C. elegans} strains.} 
Strains BY200 (vtIs1[pdat-1::GFP]) were maintained at 20$\!\!~^{\circ}\mathrm{C}$ on K-agar plates seeded with OP50 {\it E. coli}.
Strain BY200 was a gift from Michael Aschner.

% \vspace{-5pt}

\paragraph{Age-synchronization by bleaching treatment.} 
A non-starved population of day 1-2 adults was collected in a 15 $m\ell$ tube from K-agar plates by washing with K-medium.
Nematodes were allowed to settle for 2-3 minutes, and the supernatant was discarded.
Worms were then treated with 5 $m\ell$ of K-medium containing final concentrations of 0.4 $N$ sodium hydroxide and 20\% v/v sodium hypochlorite for eight minutes.
The bleaching reaction was quenched by raising the volume to 15 $m\ell$ with K-medium, centrifuged at 2200 RCF for 2 minutes, and the supernatant was discarded. 
This final washing step was repeated two more times to recover the embryos for exposure experiments. 

% \vspace{-5pt}

\paragraph{Developmental exposure to chemicals.}
Embryos generated by bleaching treatment were counted on a stereo microscope by transferring 10 $\mu \ell$ of K-medium containing embryos to a glass slide. 
K-medium was added or subtracted until reaching a concentration of 10 embryos/$\mu \ell$.
Developmental exposure was performed as described in \citep{huayta2025inhibition}.
Briefly, 100 embryos were transferred to each well of a 24-well plate, and the volume was increased to a total of 500 $\mu \ell$ per well of complete K-medium containing HB101 {\it E. coli} bacterial food at a final concentration of OD 2.0, and lead acetate (Sigma-Aldrich) or benzo[a]pyrene (Sigma-Aldrich).
The range of concentrations tested for Pb and BaP was based on previous works with these chemicals \citep{huayta2025assessment}.
All wells also contained a final concentration of 1\% v/v DMSO (dimethyl sulfoxide) as vehicle. 
The 24-well plate was put on a shaker at 20$\!\!~^{\circ}\mathrm{C}$ for 52 to 54 hours to allow nematodes to reach the L4 larval stage. The contents of each well were collected in 15 mL tubes, the total volume raised to 10 mL with K-medium, and the nematodes were allowed to settle. After 2 to 3 minutes, the supernatant was discarded, and this washing step was repeated two more times. At this point, nematodes were ready for microscopy.

% \vspace{-5pt}

\paragraph{Fluorescence microscopy.}
L4 larval-stage nematodes collected after developmental exposure were transferred to 5 $\mu \ell$ of 100 $m M$ sodium azide solution on a 2\% agarose pad on top of a glass slide to paralyze them.
After one minute, the agarose pad was covered with a coverslip. 
The glass slide was mounted on a Keyence BZ-X710 microscope equipped with a Keyence BZ-X700E metal halide light source.
Z-stacks of individual heads were acquired using a Nikon 40X objective and a Chroma GFP filter cube with 150 milliseconds of exposure, the microscope objective’s pitch of 0.5 $\mu m$, and 3$\times$3 binning.

% \vspace{-5pt}

\paragraph{Neuronal damage quantification.}
Maximum projections of the z-stacks based on maximum intensity were generated, and each dendrite of the cephalic neurons was scored as previously described in \citet{bijwadia2021quantifying}.
For dopaminergic neurons: 0 – no damage, 1 – irregular (curves), 2 – less than 5 blebs, 3 – 5 to 10 blebs, 4 – more than 10 blebs and/or breaks, 5 – breaks, 25 to 75\% dendrite loss, and 6 – breaks, more than 75\% dendrite loss.

\section{Constructing the Targeted Ray Design}\label{app_ray_design}

\begin{figure}[ht!]
    \centering
    \includegraphics[width=\linewidth, trim=120 200 130 250, clip]{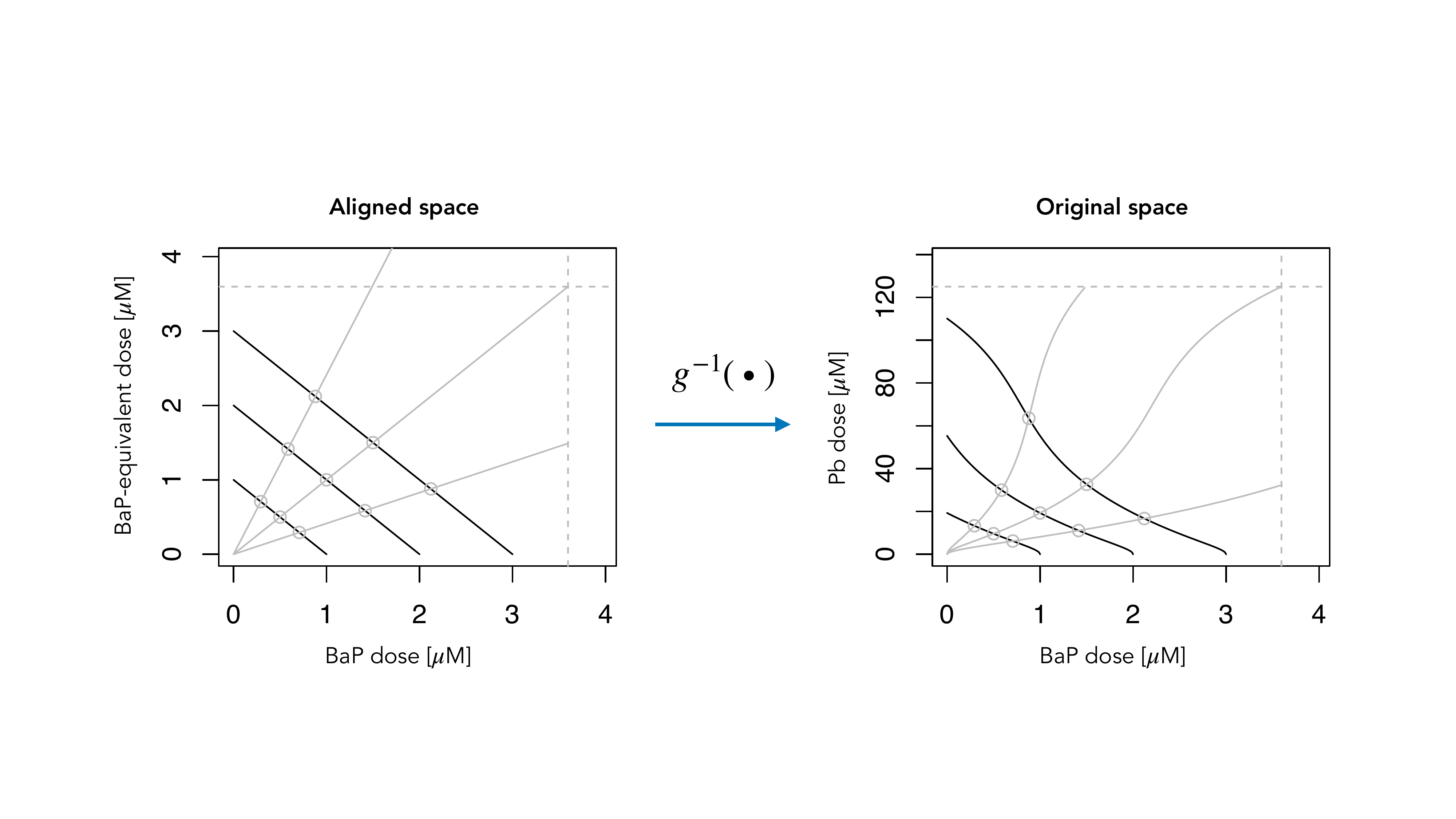}
    \put(-450,170){\makebox(0,0){\textbf{A}}}
    \put(-185,170){\makebox(0,0){\textbf{B}}}
    \vspace{-20pt}
    \caption{
    Construction of the targeted ray design from the first data batch. 
    (A) Uniform ray design in the aligned dose space, where Pb doses are rescaled to BaP-equivalent units via the alignment function $g(\cdot)$ estimated from the single-exposure measurements of the first data batch.
    Rays of fixed mixing proportion (gray lines) are spaced to cover the dose-response surface uniformly, with three non-trivial proportions $\nu \in \{1,2,3\} \cdot \pi / 8 $ chosen to span effect levels roughly equivalent to 1, 2, 3 $\mu$M of BaP alone; open circles mark the targeted dose combinations. (B) The same combinations mapped back to the original BaP–Pb dose space via $g^{-1}(\cdot)$, where the nonlinear alignment bends the evenly spaced rays of panel A into the curved trajectories actually probed in the second batch.
    }
    \label{fig_design_gInv}
\end{figure}

In this section, we provide the details on how the targeted ray design depicted in Figure~\ref{fig_data_ray_design} was chosen.
Recall that our assay began with measurements of single-toxicant exposures only, which we used to construct a principled choice of dose combinations to probe in the second stage.
Many existing studies rely on simple regular grids or ray designs.
Some contributions \citep{HollandLetz2018_OptDesigns,Sperrin2015_designs,Schurmeyer2025_OptimalDesigns} focus on optimal sample allocation among pre-specified discrete combinations within such designs -- leveraging the asymptotic variance of parametric models
 and relating it to the width of iso-effect curves or to the
combination index evaluated at pre-specified effect levels.
\citet{Huang2021_CompromiseDesign} argued that a design that is geometrically uniform in $\{d_1,d_2\}$ -- i.e. the individual doses of the two chemicals -- is optimal for surface fitting, while a design that is geometrically uniform in $\{\rho=d_1+d_2, \nu = d_1/(d_1+d_2)\}$ -- i.e. unaligned total dose and mixing proportion, as in the Hand model -- is instead optimal for testing synergy.

Targeting uniformity in the unaligned $\{\rho, \nu\}$-space
parametrization presupposes that the two doses are on a comparable scale, which is not the case here, given the substantial difference in the concentration ranges and maximum observed damage of BaP and Pb.
Hence, we modify the uniform ray design approach as follows.
Rather than targeting geometrical uniformity in the original dose space, we construct a uniform ray design in the transformed, dose-aligned space, having first estimated the alignment function $g(\cdot)$ from the single-exposure data of the first batch.
Besides the pure single-exposure rays (i.e. along the axes), we target three non-trivial mixing proportions $\nu \in \{1,2,3\} \cdot \pi / 8 $ in this aligned dose space, selected to span the dose-response surface at effect levels roughly equivalent to $\{1,2,3\}$ $\mu$M of BaP only.
We then apply $g^{-1}(\cdot)$ to map the resulting aligned dose combinations back into the original BaP--Pb dose space.
The construction process is illustrated in Figure~\ref{fig_design_gInv}, while details of the resulting dose combinations and associated sample sizes are reported in Table~\ref{tab_dose_combos_N}.

\setcounter{savedfigure}{\value{figure}}
\renewcommand{\figurename}{Table}
\addtocounter{table}{0}
\setcounter{figure}{\value{table}}
\begin{figure}[ht!]
\centering
\textbf{First Data Batch} \\[5pt]
{\setlength{\tabcolsep}{3.5pt}
\begin{tabular}{ccc|ccc}
\toprule
\multicolumn{3}{c|}{Ray Pb \textcolor{amber}{\scalebox{1.5}{$\bullet$}}} &
\multicolumn{3}{c}{Ray BaP \textcolor{indigo}{\scalebox{1.5}{$\bullet$}}} \\ 
\cmidrule{1-3} \cmidrule{4-6}
BaP & Pb & $N_{d_1d_2}^{(r)}$ & BaP & Pb & $N_{d_1d_2}^{(r)}$\\
\midrule 		
0.0 & 0 & 220 & 0.0 & 0 & 288 \\
0.0 & 5 & 247 & 0.5 & 0 & 328 \\
0.0 & 10 & 252 & 1.0 & 0 & 200 \\
0.0 & 50 & 240 & 2.5 & 0 & 237 \\
0.0 & 100 & 236 & 5.0 & 0 & 228 \\
0.0 & 125 & 255 & 10.0 & 0 & 232 \\
\bottomrule
\end{tabular}
~\\[10pt]
\textbf{Second Data Batch} \\[5pt]
\begin{tabular}{cc c| cc c| cc c| cc c| cc c}
\toprule
\multicolumn{3}{c|}{Ray Pb \textcolor{cyan}{\scalebox{1.5}{$\bullet$}}} &
\multicolumn{3}{c|}{Ray 1 \textcolor{dodgerblue}{\scalebox{1.5}{$\bullet$}}} &
\multicolumn{3}{c|}{Ray 2 \textcolor{kellygreen}{\scalebox{1.5}{$\bullet$}}} & 
\multicolumn{3}{c|}{Ray 3 \textcolor{jasper}{\scalebox{1.5}{$\bullet$}}} & 
\multicolumn{3}{c}{Ray BaP \textcolor{fandango}{\scalebox{1.5}{$\bullet$}}} \\ 
\cmidrule{1-3} \cmidrule{4-6} \cmidrule{7-9} \cmidrule{10-12} \cmidrule{13-15}
BaP & Pb & $N_{d_1d_2}^{(r)}$ & BaP & Pb & $N_{d_1d_2}^{(r)}$ & BaP & Pb & $N_{d_1d_2}^{(r)}$ & BaP & Pb & $N_{d_1d_2}^{(r)}$ & BaP & Pb & $N_{d_1d_2}^{(r)}$ \\
\midrule													
0.0 & 0 & 265 & 0.0 & 0 & 240 & 0.0 & 0 & 248 & 0.0 & 0 & 240 & 0.0 & 0 & 240 \\
0.0 & 19 & 240 & 0.3 & 13 & 244 & 0.5 &  9 & 252 & 0.7 &  6 & 239 & 1.0 & 0 & 200 \\
0.0 & 55 & 241 & 0.6 & 30 & 258 & 1.0 & 19 & 236 & 1.4 & 11 & 232 & 2.0 & 0 & 192 \\
0.0 & 110 & 235 & 0.9 & 63 & 206 & 1.5 & 32 & 220 & 2.1 & 17 & 228 & 3.0 & 0 & 139 \\
\bottomrule
\end{tabular}
}
\caption{Targeted dose combinations (in $\mu$M) for each ray in the second-stage assay. Rays Pb and BaP correspond to single-chemical exposures; 
Rays 1--3 correspond to mixture proportions $\nu \in \{1,2,3\} \cdot \pi / 8 $ in the aligned dose space.}
\label{tab_dose_combos_N}
\end{figure}
\renewcommand{\figurename}{Fig.}
\setcounter{figure}{\value{savedfigure}}

\section{Adjusting for Batch Effects via Pre-processing}\label{app_pre_process}

In this section, we provide the mathematical details of the pre-processing routine used to correct for batch effects across data batches, corresponding to rays $r=1,\dots,R$ in the targeted design shown in Figure~\ref{fig_data_ray_design}.
The key observation is that the observed counts 
\begin{equation*}
\begin{aligned}
n_{0\,\ell}^{(r)} &= {\sum_{i=1}^{N}}
\mathbbm{1} 
\big(y_i = \ell \mid d_{1i}=0,\;d_{2i}=0,\; r_i=r \big)
\hspace{55pt}
N_{0}^{(r)} = \displaystyle{\sum_{l=0}^L} n_{0\,\ell}^{(r)}
\\
n_{d_1d_2\,\ell}^{(r)} &= {\sum_{i=1}^{N}}
\mathbbm{1}
\big(y_i = \ell \mid d_{1i}=d_1,\;d_{2i}=d_2,\; r_i=r \big)
\hspace{40pt}
N_{d_1d_2}^{(r)}= \displaystyle{\sum_{l=0}^L} {n}_{d_1d_2\,\ell}^{(r)}
\end{aligned}
\end{equation*}
are sufficient statistics for our modeling approach.
Adjusting such counts for batch effects thus achieves the desired correction.
Note that here we slightly adjusted the notation used so far, distinguishing between counts for control and non-control groups.

To this end, we propose the following pre-processing routine.
We compute empirical CDFs in each control group, and transform them to an unconstrained scale
\begin{equation*}
\hat{F}_{0\,\ell}^{(r)} = \displaystyle{\sum_{\ell'\leq \ell} \frac{n_{0\,\ell'}^{(r)} + \kappa}{N_{0\,r} + (L+1) \, \kappa }}
\hspace{60pt}
\hat{m}_{0\ell}^{(r)} = \Phi^{-1}\big( \hat{F}_{0\ell}^{(r)}\big) \;.
\end{equation*}
Here, $\kappa=10^{-6}$ is a small perturbation used for numerical stability.
We then fit a random effect model $\hat{m}_{0\ell}^{(r)} \approx \alpha_\ell + \delta_r$, with fixed effects for arm membership and response values.
Estimates of the associated coefficients are available in closed form as
\begin{equation*}
\hat{\alpha}_\ell = { \frac{1}{R}\sum_{r=1}^{R}} \hat{m}_{0\ell}^{(r)}
\hspace{50pt}
\hat{\delta}_r = { \frac{1}{L}\sum_{\ell=0}^{L-1}} (\hat{m}_{0\ell}^{(r)} - \hat{\alpha}_\ell) \;.
\end{equation*}
We then use these estimates to apply a correction to all empirical CDFs on the -- across all rays, and at all dose levels -- and recompute the adjusted counts accordingly
\begin{equation*}
\begin{aligned}
\tilde{F}_{0\,\ell}^{(r)} &= \Phi\big( \hat{\alpha}_\ell 
\big)
\hspace{124pt}
\tilde{n}_{0\,\ell}^{(r)} = \big( \tilde{F}_{0\,\ell}^{(r)} - \tilde{F}_{0\,\ell-1}^{(r)}\big) \, N_{0}^{(r)} \\
\tilde{F}_{d_1d_2\,\ell}^{(r)} &= \Phi\Big(\Phi^{-1}\big( \hat{F}_{d_1d_2\,\ell}^{(r)}\big) - \hat{\delta}_r\Big)
\hspace{40pt}
\tilde{n}_{d_1d_2\,\ell}^{(r)} = \big( \tilde{F}_{d_1d_2\,\ell}^{(r)} - \tilde{F}_{d_1d_2\,\ell-1}^{(r)}\big) \, N_{d_1d_2}^{(r)} \;.
\end{aligned}
\end{equation*}
The analysis then proceeds using $\tilde{n}_{0\,\ell}^{(r)}$ and $\tilde{n}_{d_1d_2\,\ell}^{(r)}$ in place of the orginal observed counts ${n}_{0\,\ell}^{(r)}$ and ${n}_{d_1d_2\,\ell}^{(r)}$.

\vspace{-10pt}

\textcolor{black}{\section{Simplified Numerical Evaluation of the Hand Model }\label{app_hand}}

This appendix details the numerical evaluation of the Hand model.
Its defining ODE reduces to a one-dimensional integral, whose structure is further simplified by leveraging the dose-alignment function $g(\cdot)$ defined in Section~\ref{sec_method}.
%%%%%%%%%%%%%
The ODE for $\psi_{12,\nu}$ is autonomous: its right-hand side depends on the current effect level $\eta = \psi_{12,\nu}(\rho)$ only, not on $\rho=d_1+d_2$ explicitly. 
By the inverse function rule, its left-hand side can be rewritten as
$\psi'_{12,\nu} \big(\psi_{12,\nu}^{-1}(\eta)\big) =\big({\frac{\partial}{\partial \eta} \psi_{12,\nu}^{-1}(\eta)} \big)^{-1}$, thus giving
\begin{equation*}
    {\frac{\partial}{\partial \eta} \psi_{12,\nu}^{-1}(\eta)} = \frac{1}{ \nu\, \lambda_1'\big(\lambda_1^{-1}(\eta)\big) + (1-\nu)\, \lambda_2'\big(\lambda_2^{-1}(\eta)\big)} \;.
\end{equation*}
Direct integration then gives
\begin{equation}\label{eq_hand_integral}
     \psi_{12,\nu}^{-1}(\eta) = \int_0^\eta \frac{1}{ \nu\, \lambda_1'\big(\lambda_1^{-1}(\mu)\big) + (1-\nu)\, \lambda_2'\big(\lambda_2^{-1}(\mu)\big)} \, d\mu \;.
\end{equation}
Numerical evaluation is further simplified by leveraging the  dose alignment structure
$\lambda_2(d_2) = \lambda_1\big(D_1^{(\max)}\,\tau_2\,g(d_2)\big)$
used throughout our construction.
Combining
\begin{equation*}
\left\{ \,
\begin{aligned}
    \lambda_2'(d_2) &= \lambda_1' \big(D_1^{(\max)} \, \tau_2 \, g(d_2) \big) \, D_1^{(\max)} \, \tau_2 \, g'(d_2) \\[5pt]
    \lambda_2^{-1}(\mu) &= g^{-1} \big(\lambda_1^{-1}(\mu) / ( D_1^{(\max)} \, \tau_2) \big)
\end{aligned}    
\right.
\end{equation*}
we get
\begin{equation*}
    \lambda_2'\big(\lambda_2^{-1}(\mu)\big) = \lambda_1'\big(\lambda_1^{-1}(\mu)\big)\cdot D_1^{(\max)}\,\tau_2\,g'\Big(g^{-1}\big(\lambda_1^{-1}(\mu)/(D_1^{(\max)}\,\tau_2)\big)\Big)\; .
\end{equation*}
Defining $a := \lambda_1^{-1}(\mu)$,
so that $d\mu = \lambda_1'(a)\,da$, the factor $\lambda_1'(a)$ cancels between numerator and denominator, and equation~\eqref{eq_hand_integral} becomes
\begin{equation*}
    \psi_{12,\nu}^{-1}(\eta) = \int_0^{\,\lambda_1^{-1}(\eta)} \frac{da}{\nu + (1-\nu)\,D_1^{(\max)}\,\tau_2\,g'\Big(g^{-1}\big(a/(D_1^{(\max)}\,\tau_2)\big)\Big)} \;.
\end{equation*}
The integrand now depends only on $g'$ and $g^{-1}$, dispensing with $\lambda_1^{-1},\lambda_2^{-1}$ and $\lambda_2'$ altogether.
A final change of variable makes the integrand entirely forward-evaluable.
First, let us re-parametrize the effect $\eta =\lambda_2(s)=\lambda_1(D_1^{(\max)}\,\tau_2\,g(s))$ in terms of a Pb-equivalent free coordinate $s \in [0, D_2^{(\max)}]$.
Then, the change of variable $\sigma = g^{-1}\big(a/(D_1^{(\max)}\,\tau_2)\big)$ gives 
$da = D_1^{(\max)}\,\tau_2\,g'(\sigma) \, d\sigma$ and
\begin{equation}\label{eq_hand_simplified}
    \psi_{12,\nu}^{-1}\big(\lambda_2(s)\big) = \int_0^{s} \frac{D_1^{(\max)}\,\tau_2\,g'(\sigma)}{\nu + (1-\nu)\,D_1^{(\max)}\,\tau_2\,g'(\sigma)}\,d\sigma \;.
\end{equation}
Notably, this requires no root-finding for inverse function evaluation.

% \paragraph{Domain and boundary behavior.}
Recall that the original ODE is defined only where both inverse maps $\lambda_1^{-1}$ and $\lambda_2^{-1}$ exist. The BaP term requires $\eta \le 1$, but the Pb term requires $\eta \leq \eta_2^{(\max)} = \lambda_2(D_2^{(\max)})$.
Our re-parametrization $\eta =\lambda_2(s)$ is convenient here: letting $s$ vary in $[0, D_2^{(\max)}]$ automatically confines the effect grid to $\eta \leq \eta_2^{(\max)}$, so the integral is never evaluated outside the region where the Hand null is defined.
Equivalently, the combined effect along any mixed ray is undefined beyond the total dose $\rho^{(\max)}_\nu = \psi_{12,\nu}^{-1}(\eta_2^{(\max)})$, and dose pairs with
$d_1+d_2 > \rho^{(\max)}_\nu $ fall outside its domain.

% \paragraph{Implementation.}
The only numerical inversion left to perform is the one to obtain the forward combined effect $\lambda_{12}^{(o)}(d_1,d_2) = \psi_{12,\,d_1/(d_1+d_2)}(d_1+d_2)$.
For each fixed-ratio ray $\nu$, we evaluate the integral from equation~\eqref{eq_hand_simplified} on a regular grid of effects, tabulating pairs
$\Big\{ \Big( \eta = \lambda_2(s), \,
\rho = \psi_{12,\nu}^{-1}\big(\lambda_2(s)\big)
\Big) \Big\}_{s \in [0,D_2^{(max)}]}$.
% ${\big\{ \big( \eta=\lambda_2(s) , \,
% \rho = \psi_{12,\nu}^{-1}(\eta)
% \big) \big\}_{\eta \in [0,\eta_2^{(\max)}]}}$ 
from $(0,0)$ up to $\big(\eta_2^{(\max)}, \rho^{(\max)}_\nu\big)$. 
The combined effect 
% $\lambda_{12}^{(o)}(d_1,d_2) = \psi_{12,\,d_1/(d_1+d_2)}(d_1+d_2)$
at any queried dose pair $(d_1,d_2)$ on that ray is then read off by monotone interpolation of $\eta$ against $\rho = d_1 + d_2$.

\vspace{5pt}

\section{Flexible Formulations: Fit Comparison and Diagnostics}\label{app_Copulas}

In this section, we report further analysis and comparative results on flexible formulations for fitting a joint dose-response surface. 
We first provide evidence of poor fit from parametric copula models, and discuss how the mathematical properties of quasi-copulas would make them better suited for the problem at hand -- yet lacking the constructive foundations needed for practical implementation.
We then show that a spline-based quasi-copula approximation, while representing a step in the right direction, still fails to achieve satisfactory reconstruction.
We conclude by providing comparative evidence of the good
reconstruction capabilities of the proposed radial spline
formulation.

For both the radial splines and copula-based weight functions, we assess descriptive power via the ability to reconstruct a given target two-dimensional surface $\lambda_{12}^{(o)}$.
Rather than fitting the observed data via the likelihood from equation~\eqref{eq_likelihood}, we directly minimize the integrated squared deviation between the target surface and the fitted one
\begin{equation*}
\begin{aligned}
    \mathcal{D}\big(\lambda_{12}, \lambda_{12}^{(o)}\big)
    & = \int_{0}^{D_2^{(\max)}}\!\!\int_{0}^{D_1^{(\max)}}
    \Big( \lambda_{12}(d_1,d_2) - \lambda_{12}^{(o)}(d_1,d_2) \Big)^2 \, dd_1\, dd_2 \\
    & \approx \sum_{(d_1,d_2)\in\mathcal{G}}
    \Big( \lambda_{12}(d_1,d_2) - \lambda_{12}^{(o)}(d_1,d_2) \Big)^2 \; ,
\end{aligned}
\end{equation*}
evaluated on a fine regular grid
$\mathcal{G} \subset [0,D_1^{(\max)}]\times[0,D_2^{(\max)}]$
that serves as the effective training set.
This allows us to bypass the constraints imposed by the limited sample size and sparse dose combinations available in our experimental design, targeting reconstruction quality at arbitrarily high resolution.

Unlike all other analyses presented so far, the target surfaces considered in this section correspond to null baselines $\lambda_{12}^{(o)}$ constructed from the preliminary single-exposure data only -- that is, with $\lambda_{1}$ and $\lambda_{2}$ fitted using only the first data chunk, as described in Appendix~\ref{app_ray_design}.
By contrast, the two-dimensional null surfaces reported in
Section~\ref{sec_results} were obtained using the fit of $\lambda_{1}$ and $\lambda_{2}$ from all data with single-chemical exposures, from both first and second data chunks.
While this translates into slightly different isoboles with respect to Figures~\ref{fig_2D_fit}  and~\ref{fig_2D_fit_MuSyC}, the conclusions drawn below hold without loss of generality.
Our analysis below is illustrative rather than exhaustive, akin to a counterexample. 
We show failure to reconstruct a relevant representative target surface, which we deem sufficient to rule out a candidate modeling approach.

% \newpage

\subsection{Parametric Copulas}

We begin by considering parametric families for the bivariate copula $\mathcal{C}$ to be used in equation~\eqref{eq_lambda_copula}.
Specifically, we report here an analysis for the following copula families:
\begin{equation*}
\begin{aligned}
    \text{[Student-}t] \quad &
    \mathcal{C}(u,v) = \mathcal{T}_{\nu,\rho}\!\left(t_\nu^{-1}(u), t_\nu^{-1}(v)\right)
    \qquad \text{with} \quad \rho \in (-1,1), \nu > 0\\[5pt]
    %%%%%%%%%%%% 
    \text{[Frank}] \quad & 
    \mathcal{C}(u,v) = -\dfrac{1}{\theta}\log\!\bigg(1 + \dfrac{(e^{-\theta u}-1)(e^{-\theta v}-1)}{e^{-\theta}-1}\bigg)
    \qquad \text{with} \quad \theta \in \mathbb{R} \\[-1pt]
    %%%%%%%%%%%%
     \text{[Clayton}] \quad & 
     \mathcal{C}(u,v) =  1 - \!\bigg( 1 - \Big(\big(1-(1-u)^{\rho}\big)^{-\delta} +
    \big(1-(1-v)^{\rho}\big)^{-\delta} - 1 \Big)^{-1/\delta}
    \bigg)^{1/\rho}
    \\[5pt]
    %%%%%%%%%%%%
    \text{[Rayleigh}] \quad &
    \mathcal{C}(u,v) = uv + \rho\, h_\alpha(u)\, h_\alpha(v)
    \qquad \text{with} \quad \rho \in (-1,1) \;.
\end{aligned}
\end{equation*}
In the Student-$t$ copula, $\mathcal{T}_{\nu,\rho}$ and $t_\nu$ denote the bivariate and univariate Student-$t$ CDF with $\nu$ degrees of freedom and correlation $\rho$, respectively, reducing to the Gaussian copula as $\nu \to \infty$.
The Clayton copula requires that $\rho \geq 1$ and $ \delta > 0$.
In the Rayleigh copula, $h_\alpha(x) = \gamma(\alpha)(1 -
e^{\alpha(x-x^2)})$ with $\gamma(\alpha)$ a normalizing constant \citep{Zachariah2024_Copula}.

\begin{figure}[hb!]
    \centering
    \includegraphics[width=\linewidth, trim=70 20 350 20, clip]{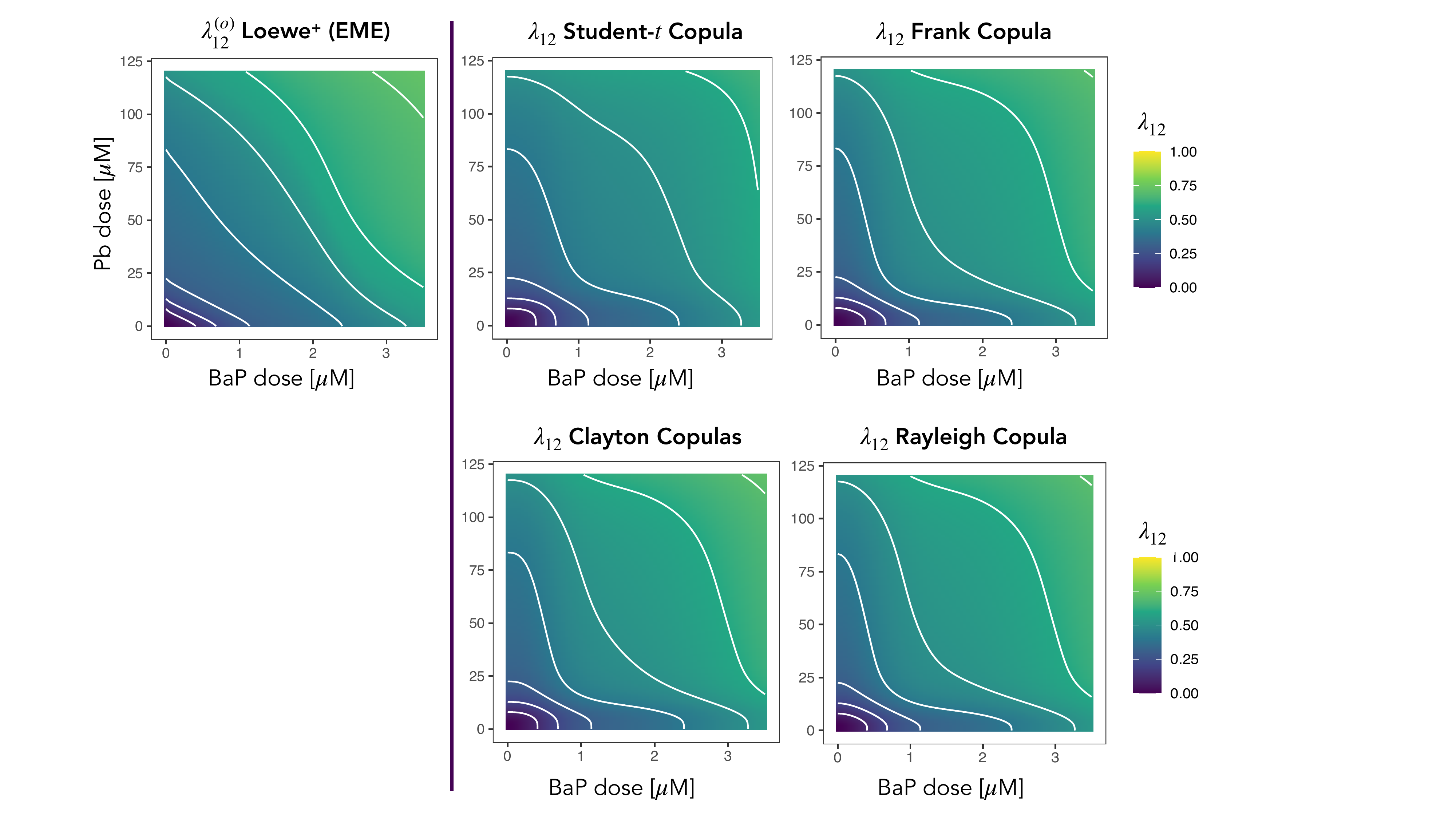}
    \vspace{-15pt}
    \caption{
    Reconstruction of the Loewe$^+$ target surface by parametric copula families.
    The leftmost panel (set off by the vertical line) shows the target additive baseline $\lambda_{12}^{(o)}$ under Loewe$^+$; the remaining panels show its best reconstructions achieved by parametric copulas -- Student-t, Frank, Clayton, and Rayleigh.
    White contours mark the isoboles, with color scales following the convention of Figure 5.
    Unlike the fits in the main text, these surfaces aim at reconstructing the target evaluated on a fine regular grid, rather than optimizing the likelihood of ordinal damage scores.
    A model reconstructs the surface well when its isoboles match those of the target in the leftmost panel, and departs from it otherwise.
    All four families fail to reproduce even the macroscopic geometry of the target, confirming that parametric copulas are insufficiently expressive for this surface.
    }
    \label{fig_lambda12_param_copula}
\end{figure}

Figure~\ref{fig_lambda12_param_copula} reports the best reconstructed dose-response weight surface achieved by each copula family, contrasted with the target $\lambda_{12}^{(o)}$ induced by Loewe$^+$.
All parametric families exhibit clear inflexibility, failing to capture even the macroscopic features of the target surface.
Parametric copulas are thus insufficiently expressive for the problem at hand.

\subsection{Spline-based Copulas and Mathematical Obstructions}

A natural next step beyond parametric families would be to model $\mathcal{C}$ via a flexible spline-based representation, in principle offering the flexibility needed to reconstruct a wide range of dependence structures.
Several nonparametric copula estimation packages are available to this end, such as \texttt{pencopula} \citep{Kauermann2013_pencopula,Schellhase2014_pencopula}, \texttt{penDvine} \cite{Schellhase2015_penDvine}, and \texttt{kdecopula} \citep{Nagler2018_kdecopula}.
These leverage B-splines, Bernstein polynomials, or kernel estimators to fit the copula density $\partial^2\mathcal{C}/\partial u\,\partial v$ from a set of empirical evaluations of its values.
In our setting, however, the copula enters as an intermediate modeling device, and no direct density evaluations are available.
More fundamentally, working at the level of the density reveals a deeper mathematical obstruction, related to the sign of the density itself.

The crux of this issue lies in a fundamental incompatibility between the mathematical properties required of any valid copula $\mathcal{C}$ and those of an admissible weight function $\lambda_{12}$.
Specifically, any dose-response weight function $\lambda_{12}$ must satisfy:
\begin{equation*}
    \begin{tabular}{r l l}
    \text{[Range]} & \textbf{1.} & $\lambda_{12} : [0,D_{1}^{(max)}] \times [0,D_{2}^{(max)}] \to [0,1]$ \\[6pt]
    \text{[Boundary Recovery]} & \textbf{2.} & $\lambda_{12}(d_1,0) = \lambda_1(d_1)$ \quad and \quad $\lambda_{12}(0,d_2) = \lambda_2(d_2)$ \\[6pt]
    \text{[Saturation]} & \textbf{3.} & $\lambda_{12}(D_{1}^{(max)},d_2) = \lambda_{12}(d_1,D_{2}^{(max)}) = 1$ \\[6pt]
    \text{[Monotonicity]} & \textbf{4.} & $\dfrac{\partial\lambda_{12}}{\partial d_1}(d_1,d_2) \geq 0$ \quad and \quad $\dfrac{\partial\lambda_{12}}{\partial d_2}(d_1,d_2) \geq 0$ \; .
    \end{tabular}
\end{equation*}
Conversley, any valid copula $\mathcal{C}$ must satisfy:
\begin{equation*}
    \begin{tabular}{r l l}
    \text{[Range]} & \textbf{1.} & $\mathcal{C} : [0,1] \times [0,1] \to [0,1]$ \\[6pt]
    \text{[Uniform Marginals]} & \textbf{2.} & $\mathcal{C}(u,1) = u$ \quad and \quad $\mathcal{C}(1,v) = v$ \\[6pt]
    \text{[Grounded]} & \textbf{3.} & $\mathcal{C}(u,0) = \mathcal{C}(0,v) = 0$ \\[6pt]
    \text{[2-increasing]} & \textbf{4.} & $\dfrac{\partial^2\mathcal{C}}{\partial u\,\partial v}(u,v) \geq 0$ \;.
    \end{tabular}
\end{equation*}
These properties are in direct correspondence with one another, with one critical exception.
The 2-increasing condition on $\mathcal{C}$ (i.e. positive density) is strictly stronger than coordinatewise monotonicity of $\lambda_{12}$.
Indeed, 2-increasing is a \emph{sufficient} condition for
monotonicity, but not a \emph{necessary} one.
This means that the class of admissible $\lambda_{12}$ is strictly larger than what can be represented through a valid copula through equation~\eqref{eq_lambda_copula}.

\begin{figure}[ht!]
    \centering
    \includegraphics[width=0.9\linewidth, trim=150 180 50 170, clip]{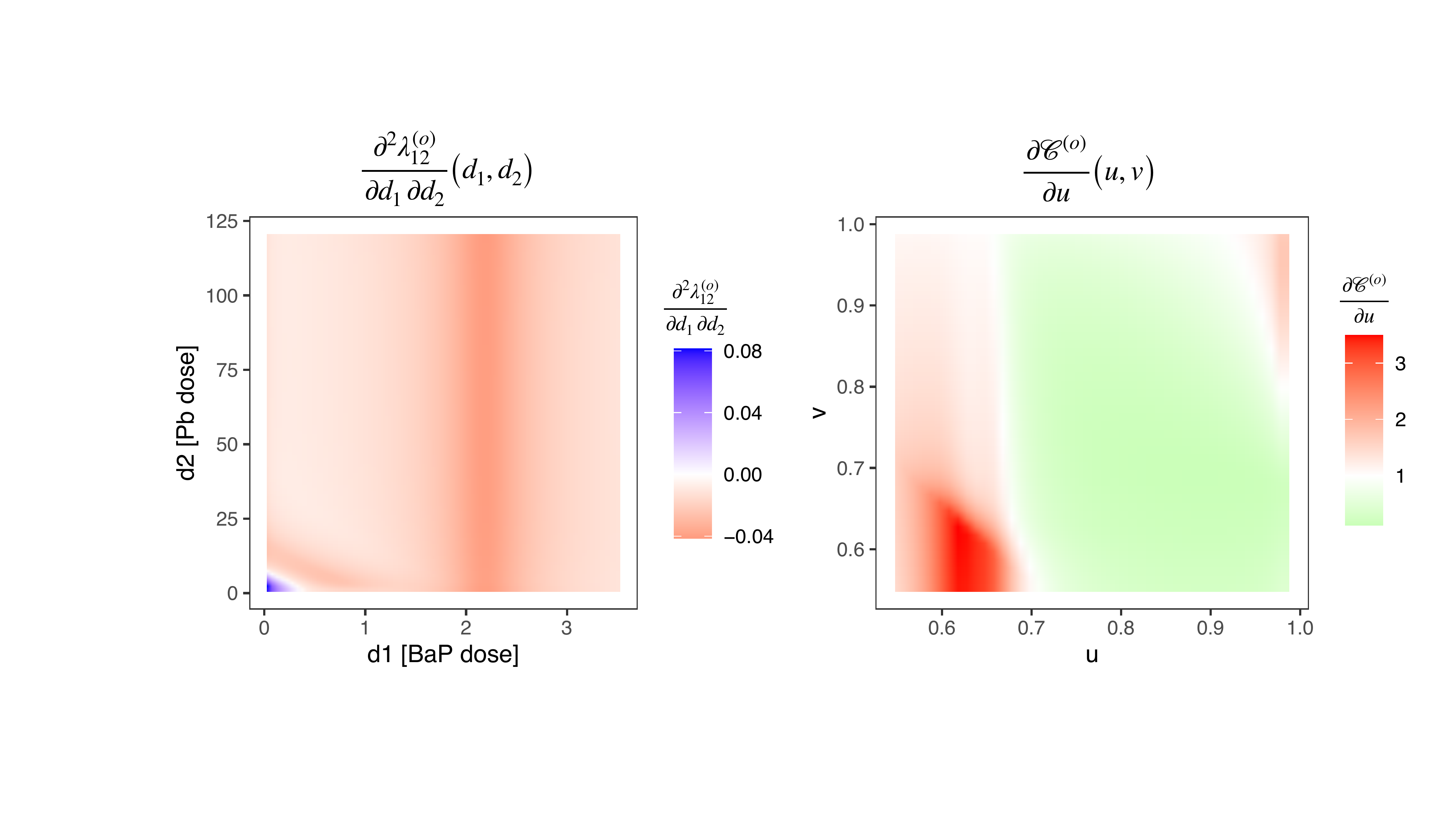}
    \put(-380,170){\makebox(0,0){\textbf{A}}}
    \put(-180,170){\makebox(0,0){\textbf{B}}}
    \vspace{-10pt}
    \caption{
    Derivatives of the Loewe$^+$ null surface $\lambda_{12}^{(o)}$ and of the induced copula-like object $Q^{(o)}$, as by equation~\eqref{eq_lambda_copula}. (A) Mixed second derivative 
    ${\partial^2 \lambda_{12}^{(o)}}/{\partial d_1 \partial d_2}$, showing that the null surface violates the 2-increasing property required of a valid copula, despite being coordinate-wise monotone by construction.
    Coherently with equation~\eqref{eq_deriv_Q_lambda}, no valid copula can thus exactly represent the EME null surface.
    (B) First derivative $\partial Q^{(o)}/\partial u$ of $Q^{(o)}$ in the $(u, v)$ space.
    Its values are bounded on the whole domain, but go over 1 on a good part of it.
    This suggests that $Q^{(o)}$ is $L$-Lipschitz but with a constant larger than the $L = 1$ typically assumed for quasi-copulas.
    }
    \label{fig_lambda12_loewe_deriv}
\end{figure}

Consider, for instance, the function $\mathcal{Q}^{(o)}$ induced by inverting equation~\eqref{eq_lambda_copula} and substituting $\lambda_{12}^{(o)}$ from the EME
\begin{equation*}
\mathcal{Q}^{(o)}(u,v)= 1 - \textstyle{\frac{1}{2}} \lambda_{1}\big(\lambda_1^{-1}(1-u) + \lambda_1^{-1} (1-v)\big) - \textstyle{\frac{1}{2}} \lambda_{2}\big(\lambda_2^{-1}(1-u) +  \lambda_2^{-1} (1-v) \big) \; .
\end{equation*}
Its density reads
\begin{equation*}
    \frac{\partial^2 \mathcal{Q}^{(o)}}{\partial u \, \partial v}\big(u,v\big)= -{\frac{1}{2}} \displaystyle{\frac{\lambda_{1}''\big(\lambda_1^{-1}(1-u) + \lambda_1^{-1} (1-v)\big) }{\lambda_1'(\lambda_1^{-1}(1-u)) \,\lambda_1'(\lambda_1^{-1}(1-v))}} -
 {\frac{1}{2}} \displaystyle{\frac{\lambda_{2}''\big(\lambda_2^{-1}(1-u) + \lambda_2^{-1} (1-v)\big)  }{\lambda_2'(\lambda_2^{-1}(1-u)) \,\lambda_2'(\lambda_2^{-1}(1-v))}}
\end{equation*}
whereas further insight comes from leveraging the first line of equation~\eqref{eq_lambda_copula} to get
\begin{equation}\label{eq_deriv_Q_lambda}
    \frac{\partial^2 \lambda_{12}^{(o)}}{\partial d_1 \, \partial d_2}\big(d_1, d_2\big) = - \lambda_1'(d_1) \, \lambda_2'(d_2) \,
    \frac{\partial^2 \mathcal{Q}^{(o)}}{\partial u \, \partial v}\big(\lambda_1(d_1), \lambda_2(d_2) \big) 
\end{equation}
Panel A of figure~\ref{fig_lambda12_loewe_deriv} displays  ${\partial^2 \lambda_{12}^{(o)}}/{\partial d_1 \partial d_2}$ under the fitted $\lambda_1$ and $\lambda_2$.
While $ \lambda_{12}^{(o)}$ is coordinatewise monotone by construction, its mixed second derivative assumes both negative and positive values over non-negligible portions of the domain.
In equation~\ref{eq_deriv_Q_lambda}, we have $\lambda_1'(d_1)\geq 0$ and $\lambda_2'(d_2)\geq 0$ by construction.
Hence, the induced $\mathcal{Q}^{(o)}$ violates one fundamental property of copulas, by having a mixed-signed density just like ${\partial^2 \lambda_{12}^{(o)}}/{\partial d_1 \partial d_2}$.
Consequently, no copula can exactly represent the EME null surface, as any copula-based approximation is structurally misaligned.

\subsection{Quasi-Copulas}

Interestingly enough, the literature offers a generalization of copulas that is more naturally aligned with our requirements.
Although less popular, quasi-copulas \citep{Nelsen2006_Copulas, Stopar2024_QuasiCopulas} relax precisely the 2-increasing condition in favor of coordinatewise monotonicity, retaining all other copula properties while adding one for smoothness:
\begin{equation*}
    \begin{tabular}{r l l}
    \text{[Range]} & \textbf{1.} & $\mathcal{Q} : [0,1] \times [0,1] \to [0,1]$ \\[6pt]
    \text{[Uniform Marginals]} & \textbf{2.} & $\mathcal{Q}(u,1) = u$ \quad and \quad $\mathcal{Q}(1,v) = v$ \\[6pt]
    \text{[Grounded]} & \textbf{3.} & $\mathcal{Q}(u,0) = \mathcal{Q}(0,v) = 0$ \\[6pt]
    \text{[Monotonicity]} & \textbf{4.} & $\dfrac{\partial\mathcal{Q}}{\partial u}(u,v) \geq 0$ \quad and \quad $\dfrac{\partial\mathcal{Q}}{\partial v}(u,v) \geq 0$ \\[6pt]
    \text{[1-Lipschitz]} & \textbf{5.} & $|\mathcal{Q}(u_2,v_2) - \mathcal{Q}(u_1,v_1)| \leq |u_2-u_1| + |v_2-v_1|$ ;.
    \end{tabular}
\end{equation*}
Note that $\partial\mathcal{Q}/\partial u \leq 1$ and
$\partial\mathcal{Q}/\partial v \leq 1$ everywhere are sufficient conditions for the 1-Lipschitz smoothness property.
Every copula is a quasi-copula, but not vice versa: quasi-copulas are strictly more general, since they induce a signed measure on $[0,1]^2$ rather than a non-negative one.
The $\mathcal{Q}^{(o)}$ induced by the EME is thus more naturally characterized as a quasi-copula than a copula, with one important caveat.
Panel B of Figure~\ref{fig_lambda12_loewe_deriv} displays
$\partial\mathcal{Q}^{(o)}/\partial u$, whose values exceed 1 over a non-negligible part of the domain.
This suggests that $\mathcal{Q}^{(o)}$ satisfies an $L$-Lipschitz condition for some $L > 1$, placing it just outside the standard quasi-copula class, which requires $L = 1$.

While extending the quasi-copula framework to accommodate $L > 1$ should be theoretically viable, quasi-copulas remain far less explored than their copula counterparts, and present additional challenges that currently limit their practical use.
This stems from the lack of a constructive characterization beyond the representation theorems, such as the one below (Theorem 10 in \citet{Dolinar2024_QuasiCopulas})
\begin{lemma}
Let $\mathcal{Q}$ be a quasi-copula. The following conditions are
equivalent:
\begin{enumerate}
\setlength{\itemsep}{10pt}
    \item[(i)] There exist copulas $\{\mathcal{C}_j\}_{j=1}^{\infty}$ and real
    numbers $\{\gamma_j\}_{j=1}^{\infty}$ such that
    \begin{equation*}
        \mathcal{Q}(x,y) = \sum_{j=1}^{\infty} \gamma_j C_j(x,y),
    \end{equation*}
    where the series converges absolutely for all $x,y \in [0,1]$.
    %%%%%%%%%%%%%%%%%%%%%%%
    \item[(ii)] The quantity 
    \begin{equation*}
        \alpha_{\mathcal{Q}} = \sup_{n \geq 1} \left\{
        \max_{i=1,\ldots,2^n} 2^n \sum_{j=1}^{2^n} V_{\mathcal{Q}}\big(R^{(n)}_{ij}\big)^+,\;
        \max_{j=1,\ldots,2^n} 2^n \sum_{i=1}^{2^n} V_{\mathcal{Q}}\big(R^{(n)}_{ij}\big)^+
        \right\},
    \end{equation*}
    is such that $\alpha_{\mathcal{Q}} < \infty$, where
    $R^{(n)}_{ij} = \left[\frac{i-1}{2^n},\frac{i}{2^n}\right] \times
    \left[\frac{j-1}{2^n},\frac{j}{2^n}\right]$, $1 \leq i,j \leq 2^n$,
    and $V_{\mathcal{Q}}(R) = \mathcal{Q}(x_2,y_2) - \mathcal{Q}(x_1,y_2)
    - \mathcal{Q}(x_2,y_1) + \mathcal{Q}(x_1,y_1)$.
    %%%%%%%%%%%%%%%%%%%%%%%%
    \item[(iii)] There exist copulas $\mathcal{A}$ and $\mathcal{B}$ and real numbers $\alpha$ and $\beta$ such that
    $$\mathcal{Q}(x,y) = \alpha A(x,y) + \beta B(x,y)$$
    for all
    $x,y \in [0,1]$.
\end{enumerate}
\end{lemma}
The last representation is the most promising, 
allowing to represent a quasi-copula as an unrestricted linear combination of two copulas.
Since a bounded density $|\partial^2\mathcal{Q}/\partial u\,\partial v| < \infty$ is a sufficient condition for $\alpha_{\mathcal{Q}} < \infty$, Panel A of figure~\ref{fig_lambda12_loewe_deriv} suggests that this representation is in principle applicable to our target $\mathcal{Q}^{(o)}$.
Note that a convex combination with $\beta = 1-\alpha \in (0,1)$ would itself be a copula rather than a non-trivial quasi-copula, so the key is allowing $\alpha$ and $\beta$ to take arbitrary real values.
However, no constructive fitting procedure currently exists for estimating $\alpha$ and $\beta$ without constraints, while simultaneously guaranteeing that the resulting linear combination satisfies all the remaining properties of a quasi-copula.

\subsection{Spline-based Quasi-Copulas}

Motivated by these limitations, we explored a more direct approach to fitting a spline-based quasi-copula-like object to the target surface.
Different contributions have pursued related strategies by relaxing copula constraints in various ways.
Most notably, \citet{Wu2012_Tensor_Splines} proposed a tensor spline-based sieve estimator for joint distributions
$\mathcal{F} = \big\{(F(s,t), F_1(s), F_2(t)) :
    (s,t) \in [L_1,U_1]\times[L_2,U_2]\big\}$
with fixed marginals and satisfying given regularity conditions.
The author proposes working in a bounded region
$[L_1,U_1]\times[L_2,U_2]$
-- with $F(L_1,L_2)>0$ and $F(U_1,U_2)<1$ --
and solving the estimation problem in a subclass of functions that increasingly approximates $\mathcal{F}$ as the sample size grows.
Crucially, this framework relaxes the requirement that $F$ be a proper joint CDF on the full domain, achieving only partial coherence for a locally accurate fit.

We adapt this idea to our setting by approximating the quasi-copula $\mathcal{Q}$ via a spline basis expansion, focusing only on the bounded region 
$[0,D_1^{(max)}]\times[0,D_2^{(max)}]$ relevant to our experimental setup, and producing a locally-consistent fit on it. 
Specifically, we approximate a quasi-copula term $\mathcal{Q}(\lambda_1(d_1),\lambda_2(d_2))$ within equation~\eqref{eq_lambda_copula} as
\begin{equation*}
    \mathcal{Q}\!\left(\lambda_1(d_1),\lambda_2(d_2)\right)
    \;\longrightarrow\;
    \rho_{12}\sum_{h=1}^{K}\sum_{h'=1}^{K}
    \theta_{hh'}\,
    \psi_h\!\left(\frac{\lambda_1(d_1)}{\lambda_1(R_1)}\right)
    \psi_{h'}\!\left(\frac{\lambda_2(d_2)}{\lambda_2(R_2)}\right),
\end{equation*}
leading to
\begin{equation*}
    \lambda_{12}(d_1,d_2) = \lambda_1(d_1) + \lambda_2(d_2) -
    \rho_{12}\sum_{h=1}^{K}\sum_{h'=1}^{K}
    \theta_{hh'}\,
    \psi_h\!\left(\frac{\lambda_1(d_1)}{\lambda_1(R_1)}\right)
    \psi_{h'}\!\left(\frac{\lambda_2(d_2)}{\lambda_2(R_2)}\right).
\end{equation*}
Here, $\{\psi_h(\cdot)\}_{h=1}^K : [0,1]\to[0,1]$ are monotone non-decreasing spline basis functions 
-- I-splines --
and $\theta \in \Delta_{K^2-1}$ are non-negative weights summing to one.

To ensure $\lambda_{12}(R_1,R_2)
\in [0,1]$, the scalar $\rho_{12}$ is constrained to satisfy 
$\lambda_1(R_1) + \lambda_2(R_2) - 1 \leq \rho_{12} \leq
\lambda_1(R_1) + \lambda_2(R_2)$.
Coordinatewise monotonicity of $\lambda_{12}$ is not automatically guaranteed and requires additional enforcement.
Analyzing the partial derivative yields the sufficient conditions
\begin{equation*}
\begin{aligned}
    & \max_{u\in(0,1)}\sum_{h}\!\Big( \sum_{h'}\theta_{hh'} \Big)\psi_h'(u)
    \leq \lambda_1(R_1)/\rho_{12}
    \\
    &
    \max_{v\in(0,1)}\sum_{h'}\!\Big( \sum_{h}\theta_{hh'} \Big)\psi_{h'}'(v)
    \leq \lambda_2(R_2)/\rho_{12}.
\end{aligned}    
\end{equation*}
Similarly to Section~\ref{sec_radial_splines}, these constraints can be enforced via a penalty term added to
the objective, augmenting the reconstruction loss with
$\gamma \, \mathcal{P}$.

Given the limited number of probed dose combinations available in practice, avoiding over-parametrization is a concrete necessity.
We therefore focus on the rank-1 parametrization
$\theta_{hh'} = \omega_{1h}\omega_{2h'}$ with
$\omega_1, \omega_2 \in \Delta_{K-1}$.
This brings the effective number of free parameters to a level comparable with that of the proposed radial spline approach, enabling a fair comparison of their
respective reconstruction power.

\begin{figure}[H]
    \centering
    \includegraphics[width=\linewidth, trim=70 550 350 20, clip]{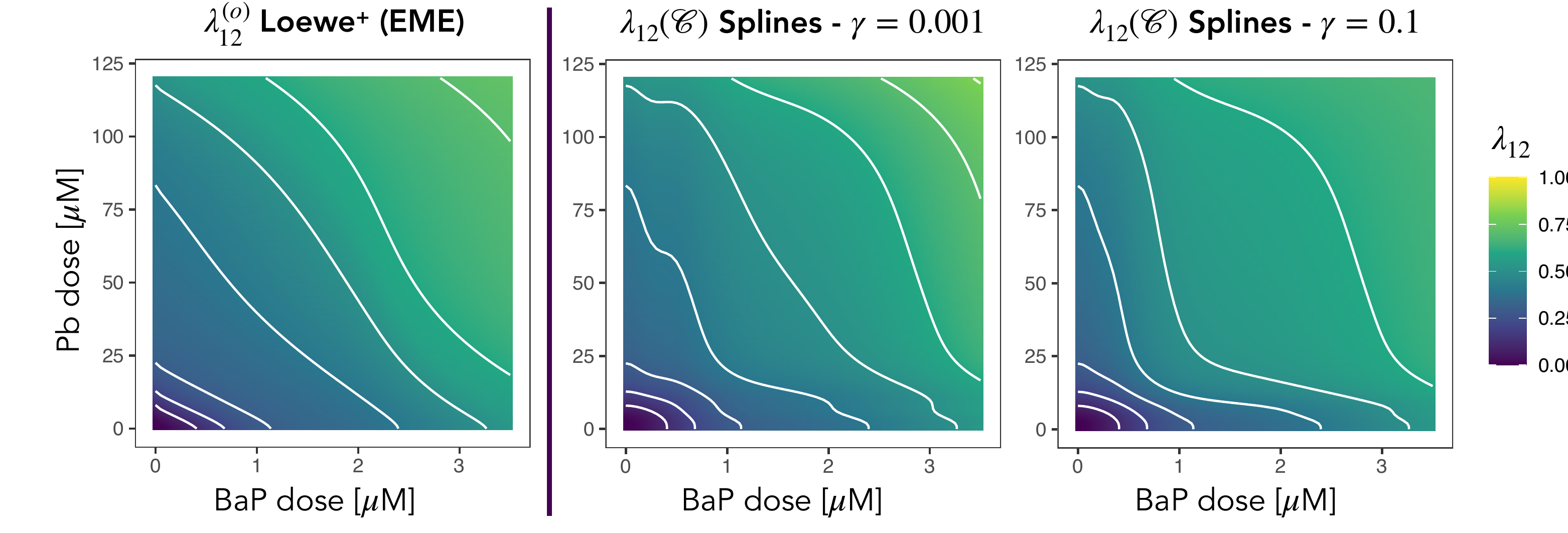}
    \vspace{-25pt}
    \caption{
    Reconstruction of the Loewe$^+$ target surface by rank-1 spline-based quasi-copulas. Leftmost panel shows the target additive baseline.
    The remaining panels show the rank-1 quasi-copula spline reconstruction at increasing penalty strength $\gamma = 0.001$ and $\gamma = 0.1$.
    % Contours and color scales follow Figure D.4, and reconstruction quality is again judged by how closely the fitted isoboles match those of the target.
    The spline-based quasi-copula improves only marginally over the parametric families, and even this gain vanishes as $\gamma$ grows, confirming that copula-style parameterizations remain insufficiently flexible for capturing realistic dose-response surfaces.
    }
    \label{fig_lambda12_splines_copula}
\end{figure}

Figure~\ref{fig_lambda12_splines_copula} reports the reconstructed surfaces under the rank-1 quasi-copula spline approach, for varying penalty strength $\gamma$.
The results suggest only minor improvements over parametric copula families, and even these minor gains vanish as $\gamma$ is increased to the point where non-negative partial derivatives are effectively enforced everywhere.
Even a spline-based quasi-copula thus lacks sufficient flexibility to reconstruct the dose-response surfaces encountered in our setting, pointing to a fundamental limitation of the copula-style parameterization that persists beyond parametric rigidity.

\subsection{Validating Flexible Splines Fit in Radial Coordinates}

\begin{figure}[ht!]
    \centering
    \includegraphics[width=\linewidth, trim=70 10 350 20, clip]{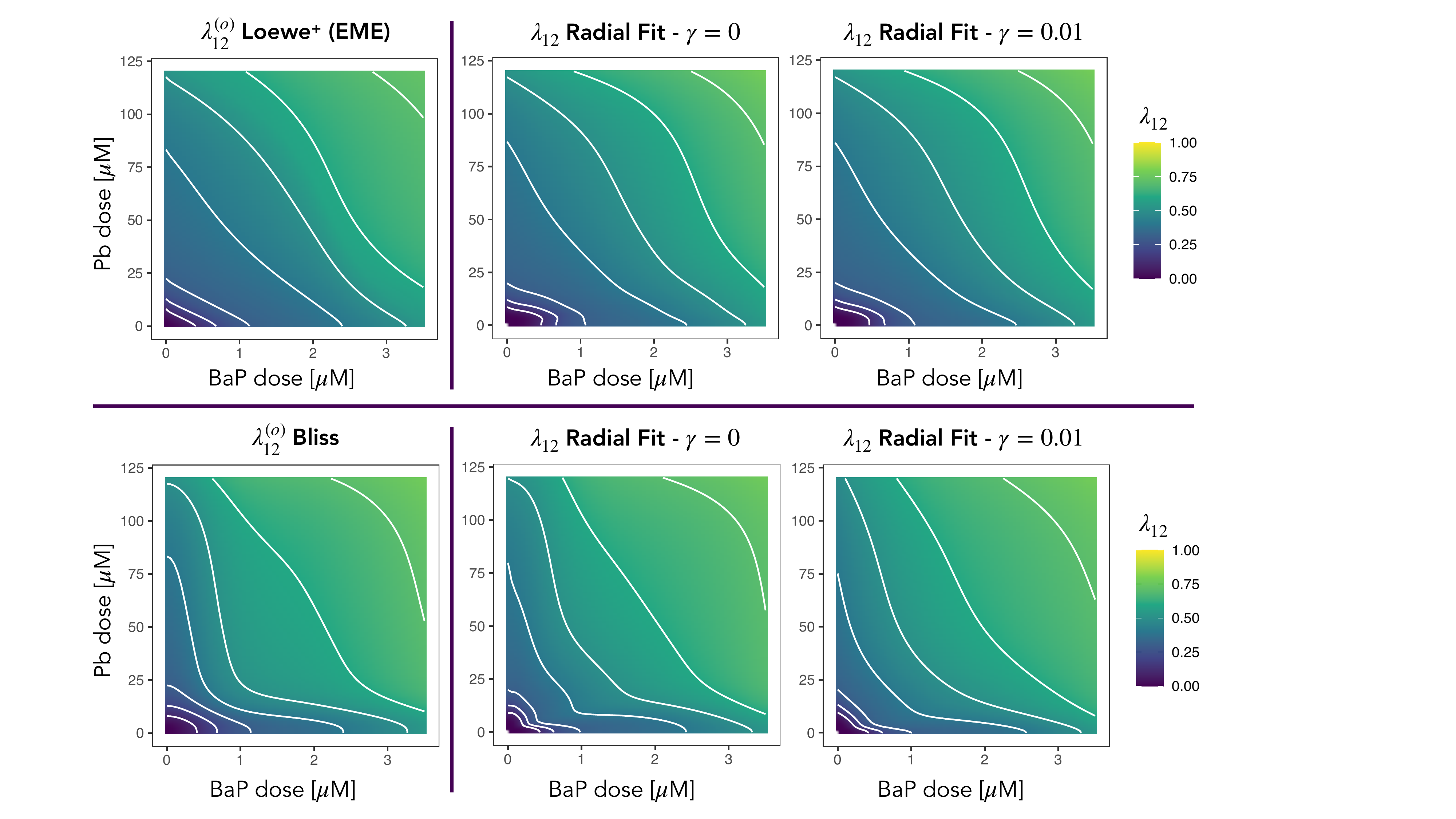}
    \vspace{-25pt}
    \caption{
    Reconstruction of Loewe$^+$ (top) and Bliss (bottom) null surfaces via the radial spline approach over the full domain.
    Leftmost column shows the target additive baselines; the remaining panels show the radial spline reconstruction of each, at penalty strength $\gamma = 0$ and $\gamma = 0.01$.
    % Contours and color scales follow Figure D.4, with reconstruction quality judged by agreement between the fitted and target isoboles.
    Unlike the copula-based approaches in Figure~\ref{fig_lambda12_param_copula} and \ref{fig_lambda12_splines_copula}, the radial spline fit closely recovers the macroscopic geometry of both null surfaces.
    }
    \label{fig_lambda12_radial}
\end{figure}

\begin{figure}[ht!]
    \centering
    \includegraphics[width=\linewidth, trim=70 10 350 20, clip]{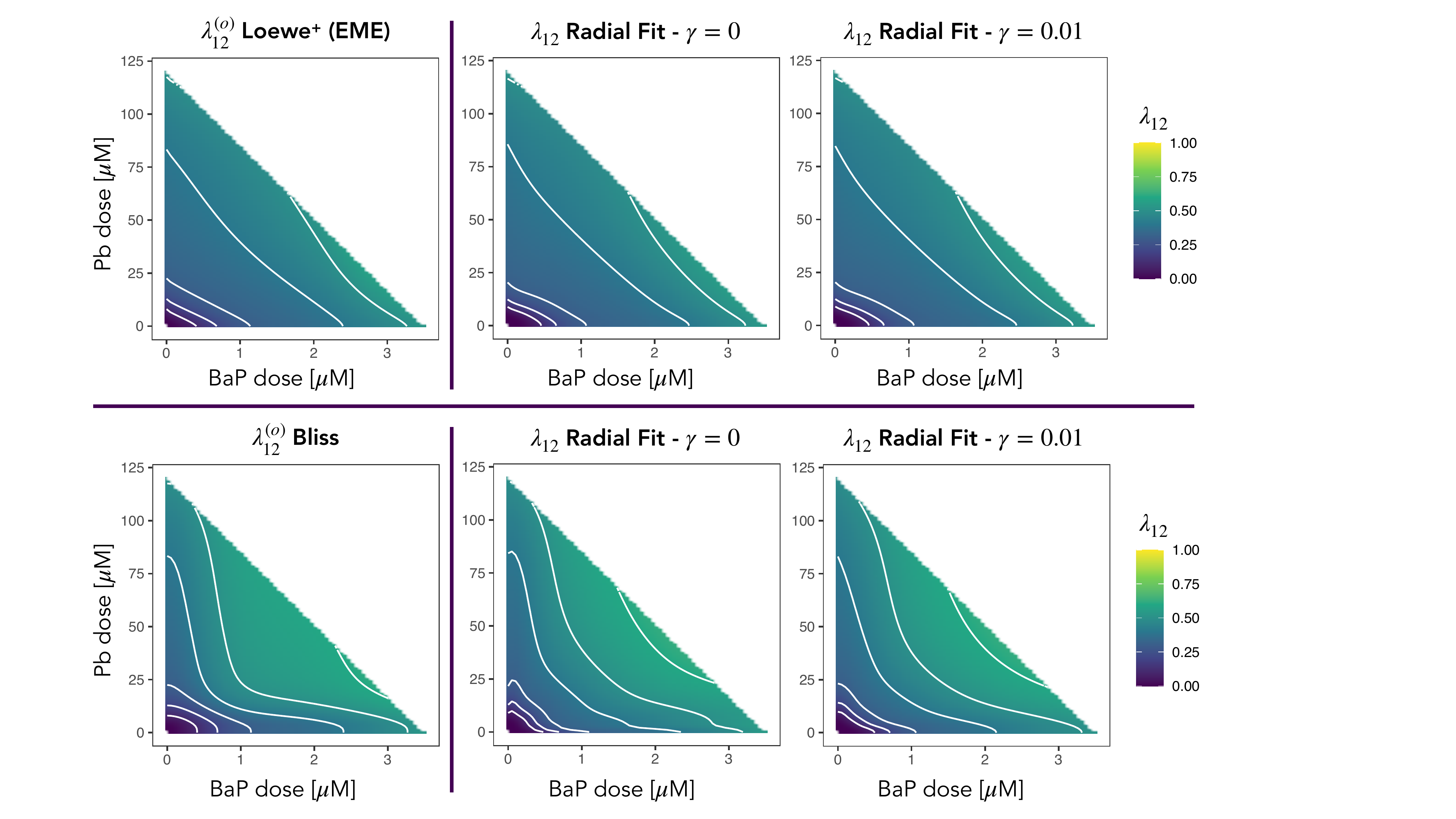}
    \vspace{-20pt}
    \caption{
    Reconstruction of Loewe$^+$ (top) and Bliss (bottom) null surfaces via the radial spline approach over the probed region. This is essentially the same analysis as in Figure D.7, but with the fit restricted to the lower-triangular dose region effectively probed in the joint-exposure experiments. Across both targets, the radial spline fit accurately reproduces the target isoboles at $\gamma = 0$ and $\gamma = 0.01$, supporting its use as the flexible $H_1$ model of choice.
    }
    \label{fig_lambda12_radial_LTR}
\end{figure}

For completeness, we apply the same analysis to the proposed radial spline approach,
assessing its ability to reconstruct the null surfaces for Loewe$^+$ and Bliss.
Figure~\ref{fig_lambda12_radial} reports the results over the full rectangular subdomain, while Figure~\ref{fig_lambda12_radial_LTR} focuses on the lower triangular region that is effectively probed in our joint exposure experiments.
Both figures show substantial improvements over the copula-based approaches presented above, with the radial spline fit accurately recovering the macroscopic features of both null surfaces across the relevant dose space.
This supports the use of our methodology as the flexible alternative of choice for the unconstrained fit under $H_1$.

\newpage

\section{Comparison with a Parametric Benchmark: MuSyC}\label{sec_MuSyC}

In this section, we report results for one of the most widely used two-dimensional parametric extensions of the Hill curve.
This contributes to the completeness of our analysis and supports the adoption of a more flexible modeling paradigm.
In particular, we focus on MuSyC \citep{Wooten2021_MuSyC}, given its claimed advantages over ZIP \citep{Yadav2015_ZIP} and BRAID \citep{Twarog2016_Braid}.
MuSyC is derived by modeling combined drug action as a reversible multi-state biochemical system at equilibrium.
Under the assumption of detailed balance, it leads to the
closed-form expression
\begin{equation*}
\lambda_{12}(d_1,d_2)=
\frac{
\omega_o +
\omega_1 \, \big( d_1/c_1 \big)^{\,h_1} +
\omega_2 \, \big( d_2/c_2 \big)^{\,h_2} +
\omega_{12} \, \alpha^{h_1} \big( d_1/c_1 \big)^{\,h_1} \big( d_2/c_2\big)^{\,h_2} 
}{ 1 +
\big( d_1/c_1 \big)^{\,h_1} +
\big( d_2/c_2 \big)^{\,h_2} +
\alpha^{h_1} \big( d_1/c_1 \big)^{\,h_1} 
}
\end{equation*}

\begin{figure}[ht!]
    \centering
    \includegraphics[width=0.9\linewidth, trim=10 50 900 20, clip]{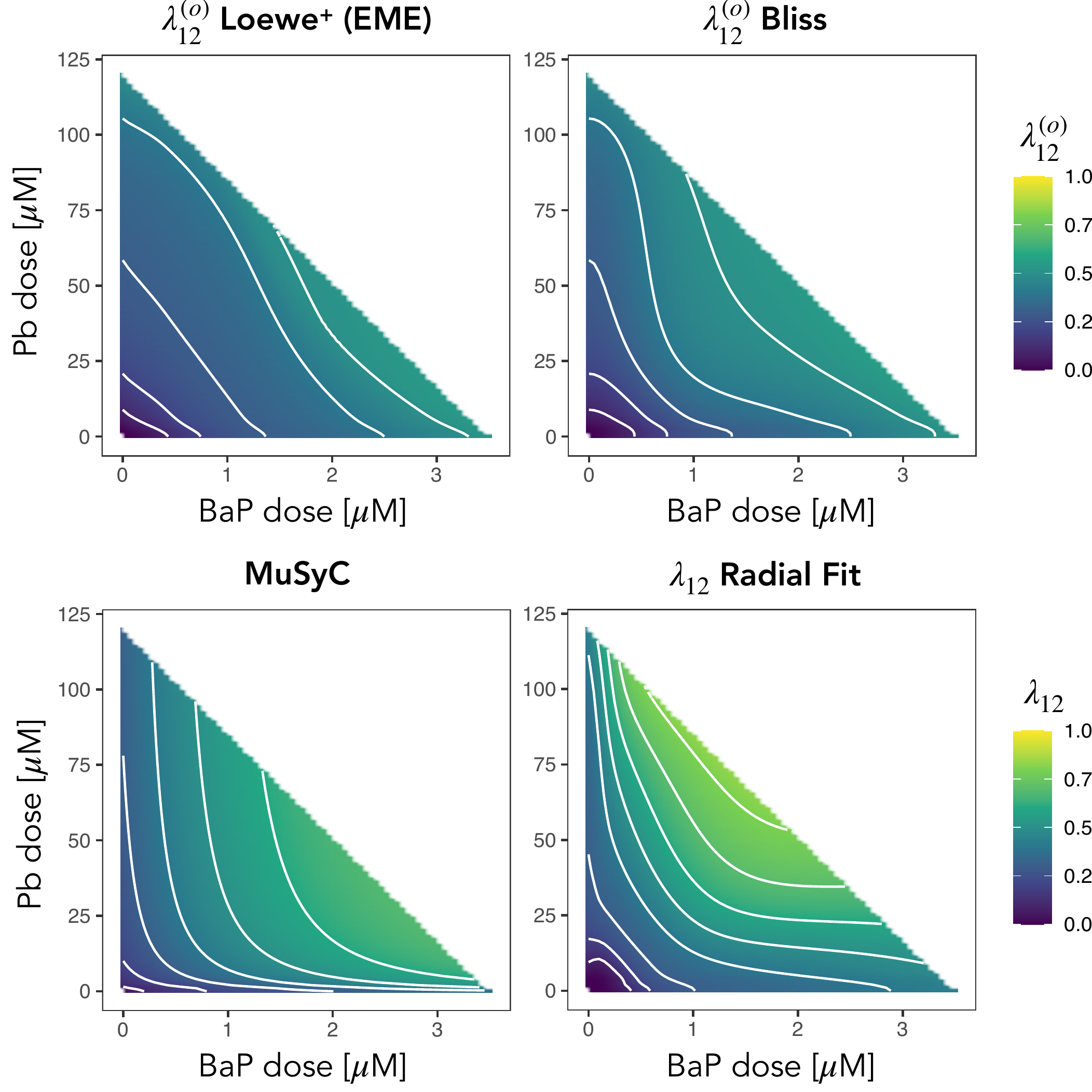}
    \put(-380,415){\makebox(0,0){\textbf{A}}}
    \put(-200,415){\makebox(0,0){\textbf{B}}}
    \put(-380,200){\makebox(0,0){\textbf{C}}}
    \put(-200,200){\makebox(0,0){\textbf{D}}}
    \vspace{-5pt}
    % \refstepcounter{section}
    \caption{
    MuSyC parametric benchmark against the additive nulls and the radial fit.
    The two null baselines $\lambda_{12}^{(o)}$ (A–B) and the radial spline fit $\lambda_{12}$ (D) are the same as in Figure 5, while the targeted-dose panel is replaced here by the MuSyC surface (C).
    Unlike the radial fit, MuSyC deviates less markedly from the additive baselines and localizes strongest departures in a different region of the dose space, concentrated at low BaP rather than in the moderate-BaP/high-Pb region flagged by the radial fit.
    Its parametric form yields visibly more rigid isobole shapes, unable to accommodate the localized synergy pattern captured by the radial spline fit.}
    \label{fig_2D_fit_MuSyC}
\end{figure}
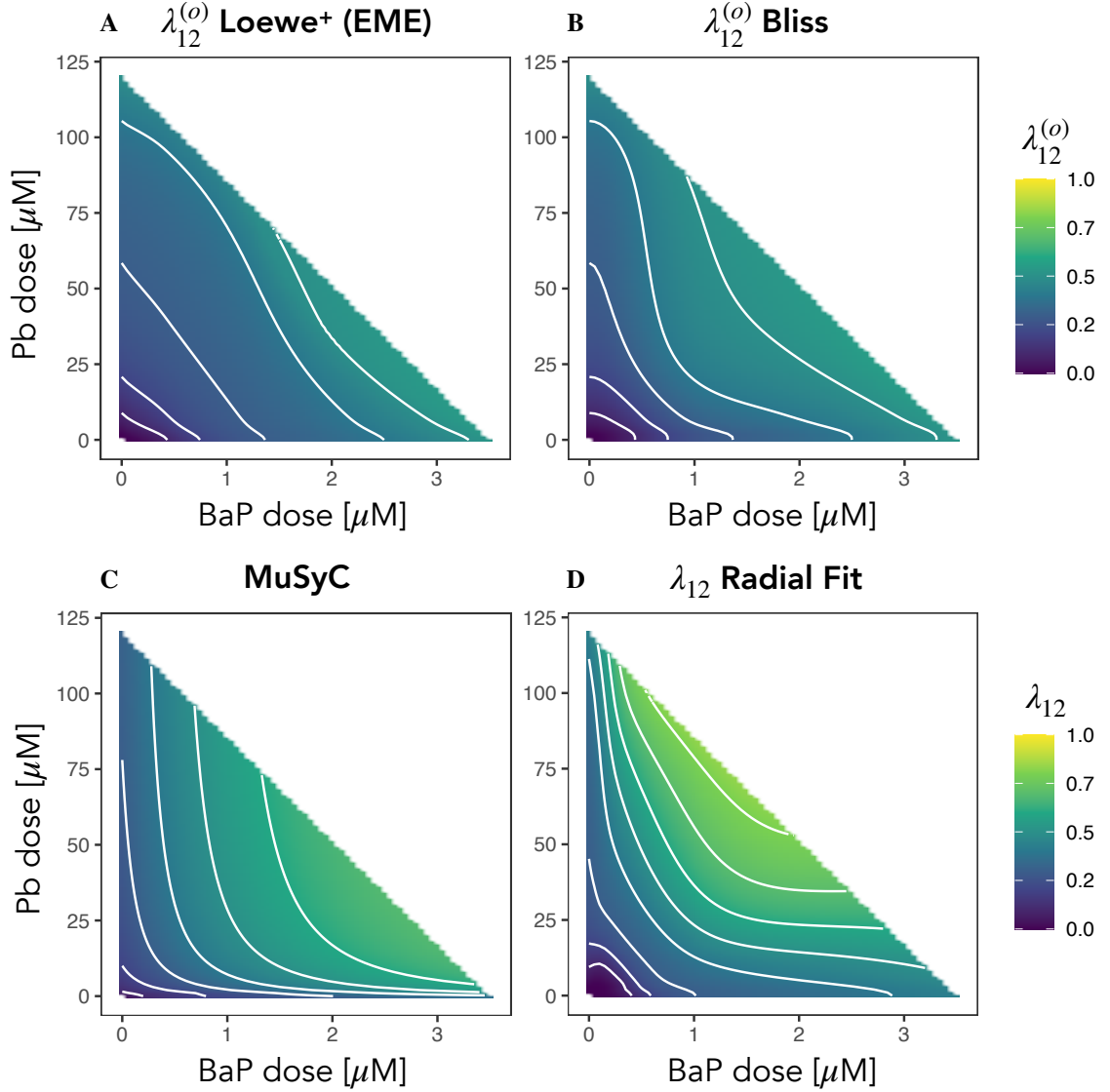

Figure~\ref{fig_2D_fit_MuSyC} reports the dose-response surface fitted by MuSyC, alongside the two additive nulls and the radial spline fit from Section~\ref{sec_results}.
Even visually, the MuSyC surface disagrees with the radial fit, showing less strong deviation from additivity and localizing the departure in a different region of the dose space.

\newpage

\begin{figure}[ht!]
    \centering
    \includegraphics[width=\linewidth, trim=170 260 120 280, clip]{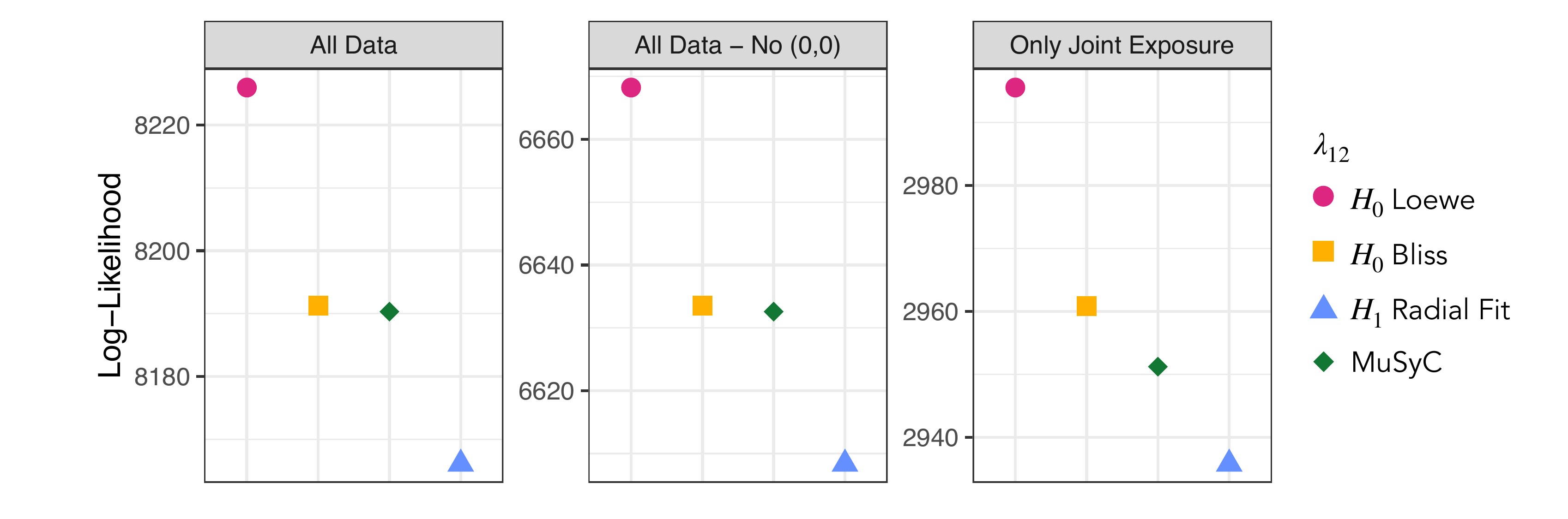}
    \put(-420,140){\makebox(0,0){\textbf{A}}}
    \put(-305,140){\makebox(0,0){\textbf{B}}}
    \put(-190,140){\makebox(0,0){\textbf{C}}}
    \\
    \includegraphics[width=\linewidth, trim=130 190 120 190, clip]{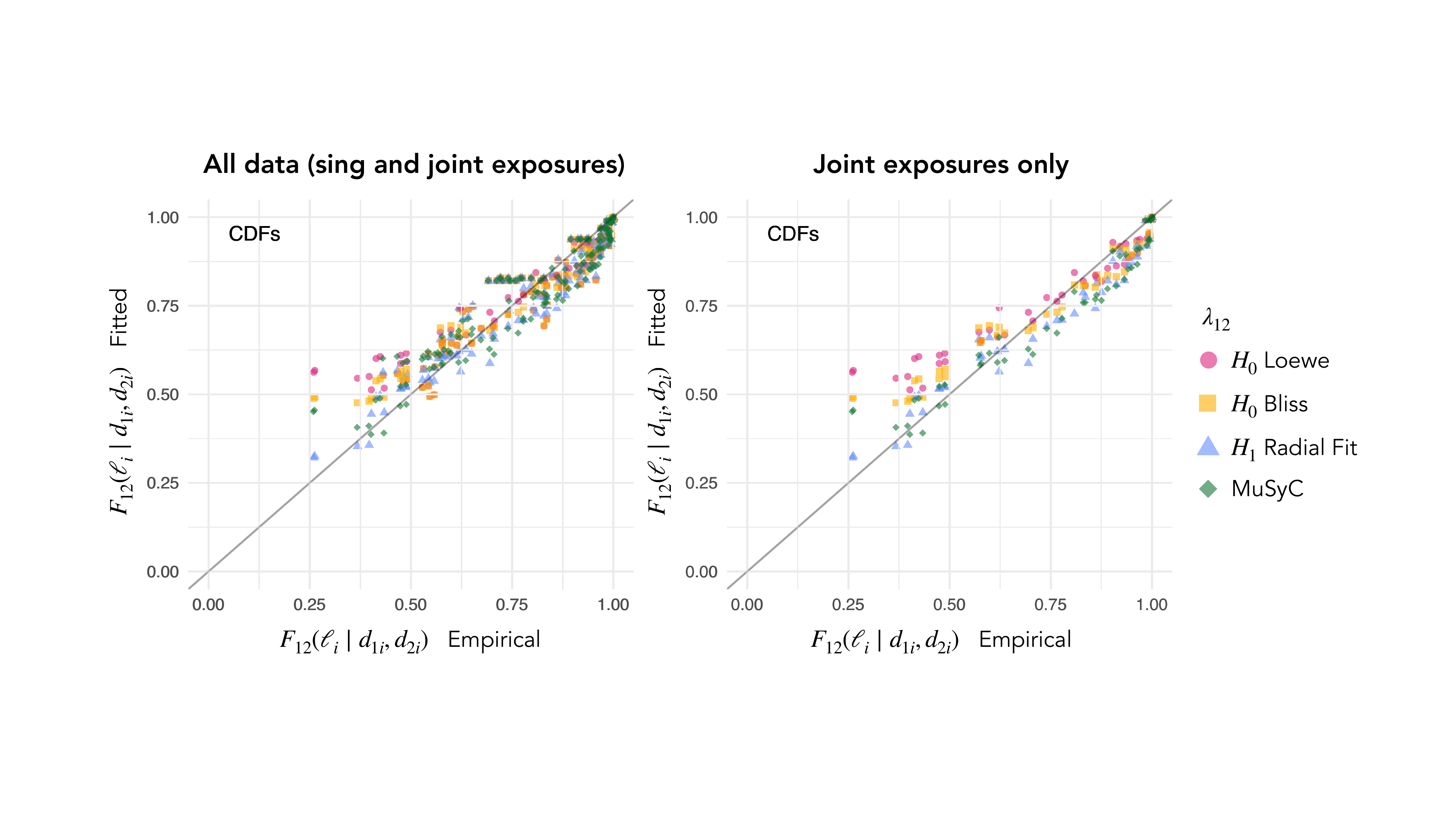}
    \put(-450,190){\makebox(0,0){\textbf{D}}}
    \put(-250,190){\makebox(0,0){\textbf{E}}}
    \put(-395,185){\colorbox{white}{\makebox(120,10){}}}
    \vspace{-20pt}
    \caption{
    Model fit comparisons including the MuSyC benchmark. 
    Panels and conventions follow Figure 7, with MuSyC added to the set of compared models.
    (A–C) Log-likelihood across data subsets: MuSyC improves only marginally over the null models and remains well short of the radial fit, notably with gaps for both single and joint-exposure data (C). 
    (D–E) Fitted versus empirical CDFs for all exposures (D) and joint exposures only (E). 
    MuSyC lies closer to the null models than the radial fit, failing to recover the empirical CDFs at the joint-exposure combinations that drive the synergy signal -- the same dose region where the radial fit achieves its sharpest improvement.
    }
    \label{fig_likelihood_MuSyC}
\end{figure}

Figure~\ref{fig_likelihood_MuSyC} further unpacks this fit by comparing log-likelihoods and CDF scatter plots across all fitted surfaces.
MuSyC improves marginally over the null models but falls
substantially short of the radial fit. 
In particular, it fails to appropriately capture the distribution of the ordinal outcomes at the dose combinations that are most critical in driving the synergy results reported in Section~\ref{sec_results}, suggesting that its parametric structure is insufficiently flexible for the dose-response profile observed in this assay.

{\section{Hypothesis Test Calibration via Bootstrap}\label{app_bootstrap}}

The likelihood ratio test of Section~\ref{sec_results} is calibrated by a parametric bootstrap to obtain the null distribution of the test statistic $T(y)$.
Specifically, in our analysis we focus on a conditional test:
we resample only the joint-exposure observations, holding the single-exposure fits $\lambda_1(\cdot)$ and $\lambda_2(\cdot)$ fixed at their estimated values -- together with the alignment $g(\cdot)$ and the scalar
$\tau_2$.
The single-chemical dose-response curves are thus treated as fixed rather than resampled, so the resulting null distribution and $p$-values are conditional on them.
%%%%%%
While this choice was mainly motivated by computational limitations, it remains statistically well-founded. 
On the computational side, even the conditional bootstrap as currently implemented was run in parallel across multiple nodes on a computing cluster to complete within a practical timeframe.
Resampling the single-toxicant data would require re-estimating the single-exposure curves -- and, with them, every null surface $\lambda_{12}^{(o)}$ -- within each of the $10{,}000$ replicates, increasing the computational cost even further.
Statistically, conditioning on the single-exposure curves is well-justified in our setting, on grounds of both relevance and magnitude.

The scientific question the test needs to address is, in substance, one of interaction between the two chemicals.
Hence, the relevant signal lies in the joint exposures alone, where the observed response is tested against the additive baseline that the single-exposure data serve to construct.
Each null $\lambda_{12}^{(o)}$ is a functional of $\lambda_1$ and $\lambda_2$, and reconstructs them along the axis.
The additivity question pertains entirely to how these are combined, not in the single-exposure curves themselves.
Besides, the fitted $\lambda_{12}$ also reconstructs the single-toxicant dose-response with high accuracy, if not exactly.
Even empirically, the fit gap between the null models and the radial fit is concentrated almost entirely in the joint-exposure data -- as by panels A-C of Figure~\ref{fig_likelihood} -- with the single-exposure profiles essentially indistinguishable across methods.

This near-agreement also bounds the uncertainty that conditioning omits.
Perturbing $\lambda_1,\lambda_2$ shifts both $\lambda_{12}^{(o)}$ and $\lambda_{12}$, but, since the two track each other on the single-chemical axes, the associated induced variability is largely common to the two terms and cancels in the difference.
This leaves $T(y)$ dominated by the discrepancy in fit on
the joint exposures, whereas the omitted residual variability represents a higher-order correction.
%%%%%
Finally, the single-exposure data are considerably more abundant than the joint-exposure ones in our assay -- as shown in Table~\ref{tab_dose_combos_N} -- so that the
single-exposure curves are estimated more precisely to begin with. 
The uncertainty left unaccounted for by conditioning on the single-exposure curves is therefore reasonably smaller than the joint-exposure sampling variability that the bootstrap does retain.

The conditional scheme thus fixes the better-determined quantities and resamples the more variable ones, propagating the dominant source of uncertainty while keeping the computation affordable.

\vspace{10pt}

\putbib
\end{bibunit}

\end{document}